\documentclass[twocolumn,showpacs,superscriptaddress,prb]{revtex4-1}    
\usepackage[utf8x]{inputenc}
\usepackage{graphicx,amsmath,mathrsfs}
\usepackage{amssymb}
\usepackage{amsthm}
\usepackage{bm}
\usepackage{subfigure,xcolor}   

\def\bc{\begin{center}}
\def\ec{\end{center}}

\newcommand{\bs}[1]{\boldsymbol{#1}}

\newcommand{\ket}[1]{\left|#1\right\rangle}

\def\ie{\emph{i.e.},\ }
\def\eg{\emph{e.g.}\ }
\def\ea{\emph{et al.\ }}

\newcommand{\pd}{{\phantom{\dag}}}
\newcommand{\up}{\uparrow}
\newcommand{\dw}{\downarrow}

\begin{document}
\title{Rashba spin orbit coupling in the Kane-Mele-Hubbard model}
\author{Manuel Laubach}
\affiliation{Institute for Theoretical Physics, University of W\"urzburg, 97074 
W\"urzburg, Germany}
\author{Johannes Reuther}
\affiliation{Department of Physics, California Institute of Technology, Pasadena, CA 91125, USA}
\author{Ronny Thomale}
\affiliation{Institute for Theoretical Physics, University of W\"urzburg, 97074 
W\"urzburg, Germany}
\author{Stephan Rachel}
\affiliation{Institute for Theoretical Physics, TU Dresden, 01062 Dresden, Germany}


\begin{abstract}
Spin-orbit (SO) coupling is the crucial parameter to drive topological insulating phases in electronic band models. In particular, the generic emergence of SO coupling involves the Rashba term which fully breaks the SU(2) spin symmetry. As soon as interactions are  taken into account, however, many theoretical studies have to content themselves with the analysis of a simplified U(1) conserving SO term without Rashba coupling. We intend to fill this gap by studying the Kane-Mele-Hubbard (KMH) model in the presence of Rashba SO coupling and present the first systematic analysis of the effect of Rashba SO coupling in a correlated two-dimensional topological insulator. We apply the variational cluster approach (VCA) to determine the interacting phase diagram by computing local density of states, magnetization, single particle spectral function, and edge states. Preceded by a detailed VCA analysis of the KMH model in the presence of U(1) conserving SO coupling, we find that the additional Rashba SO coupling drives new electronic phases such as a metallic regime and a weak topological semiconductor phase which persist in the presence of interactions.
\end{abstract}

\pacs{03.65.Vf,71.27.+a,73.20.-r}


\maketitle

\section{Introduction}

Since their theoretical prediction\,\cite{kane-05prl146802,kane-05prl226801,bernevig-06prl106802,bernevig-06s1757} and experimental discovery\,\cite{koenig-07s766}, topological insulators\,\cite{hasan-10rmp3045,qi-11rmp1057,Bernevig2013} have become one of the most vibrant fields in contemporary condensed matter physics. In two spatial dimensions, the topological insulating state can be interpreted as the spin-type companion of the charge-type integer quantum Hall effect on a lattice. For the quantum spin Hall effect, the characteristic feature to drive a given electronic band model into this topologically non-trivial phase is the band inversion due to spin-orbit (SO) coupling. As the kinetic and spin degree of freedom are coupled due to SO coupling, the electronic band structure loses its SU(2) spin symmetry. Two different types of SO coupling can be distinguished: (i) the intrinsic spin orbit coupling $V_{\text{ISO}}\sim (Z^4) L^z S^z$ where the SU(2) spin group is only broken down to U(1) (\ie retaining a conserved $S^z$ quantum number) and (ii) the Rashba SO coupling $V_{\text{RSO}}\sim \bs{E} \cdot (\bs{S} \times \bs{p})$ which does not retain a conserved continuous subgroup of SU(2). 
While the intrinsic SO coupling gives rise to the topological insulator phase, the Rashba SO coupling itself is unable to induce the non-trivial topology.
In any experimental situation, due to the presence of \eg a substrate or external electric fields, Rashba SO coupling needs to be taken into account. 

As the first microscopic model for topological insulators, the Kane-Mele model was originally proposed to describe the quantum spin Hall effect in graphene\,\cite{kane-05prl146802,kane-05prl226801}. Subsequent band structure calculations showed, however, that the spin orbit gap in graphene is so small\,\cite{min-06prb165310,yao-07prb041401} that the QSH effect  in graphene is beyond  any experimental relevance. Still, Kane and Mele's pioneering proposal for a prototypical topological insulator has triggered an intensive search for possible realizations. In principle, the spin-orbit coupling $\lambda$ can be increased using heavier elements since $V_{\rm ISO}\propto Z^4$ as a function of the atomic coordination number $Z$. Hence, promising proposals include graphene endowed with heavy adatoms like indium and thallium\,\cite{weeks-11prx021001}, synthesized silicene\,\cite{liu-11prl076802,ezawa12prl055502} (monolayers of silicon),  molecular graphene\,\cite{ghaemi-12prb201406}, honeycomb films of tin~\cite{xu-13prl136804}, monolayers or thin films of the Iridium--based honeycomb compounds X$_2$IrO$_3$ (X=Na or Li)\,\cite{shitade-09prl256403,jenderka-13prb045111}, and ``digital'' transition metal oxide heterostructures\,\cite{xiao-11nc596}. Alternatively, the Kane-Mele model might be realized using ultra-cold atoms in tunable optical lattices\,\cite{bloch-08rmp885}. Very recent progress has been made in realizing honeycomb optical lattices\,\cite{Soltan-Panahi-11np434} as well as non-Abelian gauge fields acting as a synthetic spin orbit coupling\,\cite{lin-09n628,goldman-10prl255302,dalibard-11rmp1523,lin-11n83}. Furthermore, a completely different route to realize the quantum spin Hall effect on the honeycomb lattice is to induce it by virtue of interactions\,\cite{raghu-08prl156401,lei-12prb235135,wang-12epl57001,budich-12prb201407,garcia-matrinez-13arXiv:1308.6094,daghofer-13arXiv:1308.6211,roy-13prb045425,araujo-13prb085109}.

At the non-interacting level, a Rashba SO term has already been considered in the original work by Kane and Mele  where it is shown that the QSH phase of non-interacting fermions is stable with respect to a breaking of $S_z$ symmetry. It is also argued that the otherwise quantized spin Hall conductance will deviate from its quantized value in the presence of a Rashba term\,\cite{kane-05prl146802,kane-05prl226801}. Later it was explicitly shown that the QSH phase survives the combination of disorder and Rashba spin orbit coupling but the value of the spin Hall conductance deviates significantly from the quantized value\,\cite{sheng-06prl036808}. 

For the purpose of including interactions in the Kane-Mele model, theoretical approaches have preferably constrained themselves to the exclusive consideration of intrinsic spin orbit coupling. There are two main reasons for this development. First, some theoretical approaches such as quantum Monte Carlo (QMC) necessitate the U(1) symmetry kept by the intrinsic SO coupling in order to be applicable, \ie in the case of QMC, to avoid the sign problem.
Second, calculating the topological invariant in terms of single particle Green's functions in the absence of inversion symmetry as implied by Rashba SO coupling is significantly more complicated, and often yields an integral form of the Volovik invariant\,\cite{volovik03} which is not amenable to efficient numerical evaluation. 
The Kane-Mele model with an onsite Hubbard interaction term and only intrinsic spin-orbit coupling has been usually referred to as {\it Kane-Mele-Hubbard} (KMH) model and attracted much attention recently; it was investigated from many different perspectives\,\cite{rachel-10prb075106,hohenadler-11prl100403,soriano-10prb161302,wu-12prb205102,dong-11prb205121,yamaji-11prb205122,yu-11prl010401,lee11prl166806,wen-11prb235149,mardani-11arXiv:1111.5980,hohenadler-12prb115132,griset-12prb045123,vaezi-12prb195126,assaad-13prx011015,araki-13prb205440,hung-13prb121113,ueda-13prb161108,zare-13prb224416,araki-13arXiv:1311.3973,meng-13arXiv:1310.6064,hung-13arXiv:1307.2659} providing us with a fairly good understanding of its phase diagram: For weak interactions, the topological insulator remains stable and the metallic edge states persist. For intermediate interactions, a phase transition into a magnetically ordered phase occurs. The latter has been shown to exhibit easy plane antiferromagnetic order\,\cite{rachel-10prb075106} and the transition to be of 3D $XY$ type\,\cite{hohenadler-12prb115132,wu-12prb205102}. In the isotropic limit of vanishing spin orbit coupling, one finds the semi-metallic phase (weak interactions) of graphene as well as the N\'eel antiferromagnet (strong interactions), with the phase transition of regular 3d Heisenberg type~\cite{assaad-13prx031010}. Also related correlated TI models have been studied\,\cite{yoshida-12prb125113,yoshida-13prb085134}.
(For a review about correlation effects in topological insulators see Ref.\,\onlinecite{hohenadler-13jpcm143201}.)

\begin{figure}
    \includegraphics[width=.37\textwidth]{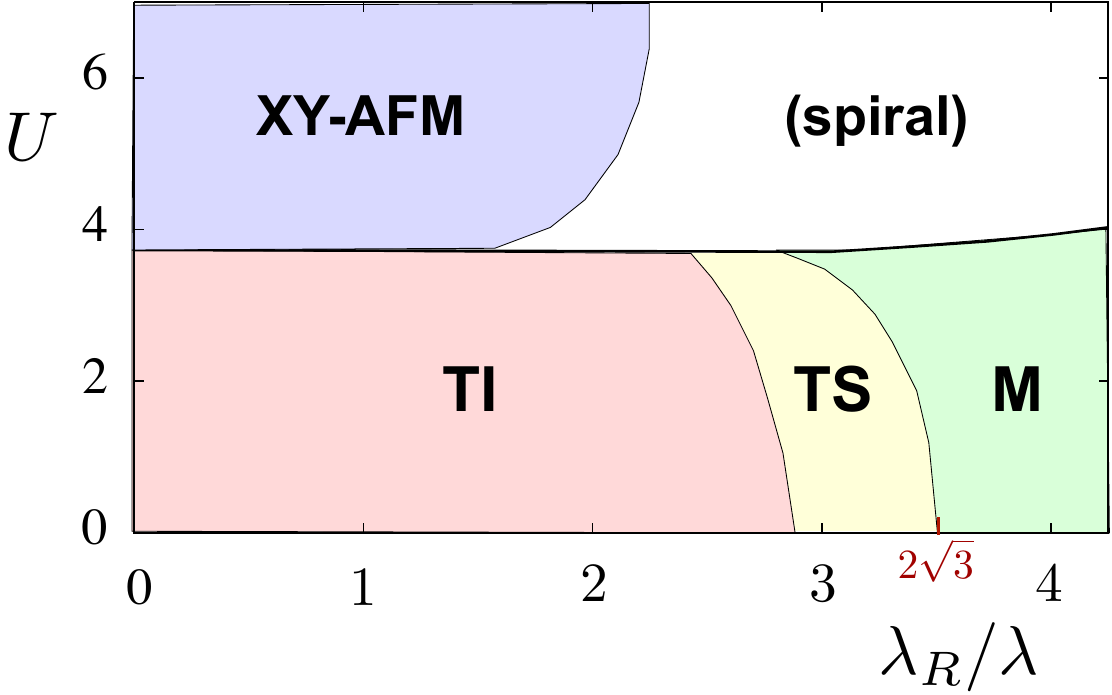}
     \caption{(Color online). Schematic $U$--$(\lambda_R/\lambda)$ phase diagram of the full Kane-Mele-Hubbard model for $\lambda = 0.2$ ($t=1$). There are five different phases: topological insulator (TI), weak topological semiconductor (TS), metal (M), easy plane antiferromagnet (XY-AFM), and possibly a phase with incommensurate spiral order. For larger $\lambda$ the TS phase becomes broader while for smaller $\lambda$ the TS phase shrinks until it vanishes for $\lambda<0.1$}
    \label{fig:schematic-phasedia}
\end{figure} 

Bridging the gap between possible experimental realizations and theoretical modeling, taking into account Rashba SO coupling and interactions in the Kane Mele model is indispensable.
We emphasize that the effect of Rashba SO coupling has so far not been investigated in any  two-dimensional correlated topological insulator model (with the exception of the one-dimensional edge theory of topological insulators dubbed {\it helical Luttinger liquid}\,\cite{wu-06prl106401,strom-10prl256804,budich-12prl086602,schmidt-12prl156402}).
In this article, we employ the variational cluster approach (VCA)\,\cite{potthoff-03prl206402,potthoff03epjb429} to investigate the generalized Kane-Mele-Hubbard model in the presence of Rashba spin orbit coupling. The VCA is an efficient method to investigate interaction effects in correlated electron systems and to obtain effective electronic band structures.  Our main results are summarized in Fig.\,\ref{fig:schematic-phasedia}. For small Rashba coupling, we find the TI (at small onsite interaction $U$) and XY-AFM phases (at large interactions $U$) which are also present in the Kane-Mele-Hubbard model without the Rashba coupling. Larger Rashba coupling induces a topologically non-trivial direct-gap only semiconductor before the system eventually becomes metallic. The XY-AFM phase is found to break down at large Rashba couplings beyond which the evolving magnetic phase cannot be analyzed anymore via VCA due to limited cluster size. Involving the knowledge from alternative approaches such as pseudofermion functional renormalization group~\cite{reuther-12prb155127,reuther-14prb100405}, this parameter regime is conjectured to be dominated by incommensurate spiral order.

The paper is organized as follows. In Sec.\ II, we introduce the Kane-Mele-Hubbard model and briefly describe the variational cluster approach (VCA). In Sec.\,III, we establish a first VCA benchmark by showing results for the KMH model in the absence of Rashba spin orbit coupling. This scenario serves as a prototypical framework to illustrate various subtle issues in the VCA approach such as cluster dependence, 
where details  are delegated to Appendix\,A. Subsequently, the results for the KMH model in the presence of finite Rashba SO coupling are presented in Sec.\,IV. In Sec.\,V, we conclude that the non-trivial phases of the Kane-Mele model emerging due to Rashba SO coupling persist in the presence of interactions, and that the interplay of interactions and Rashba SO coupling establishes a promising direction of study in theory and experiment.

\begin{figure}[t]
    \includegraphics[scale=0.4]{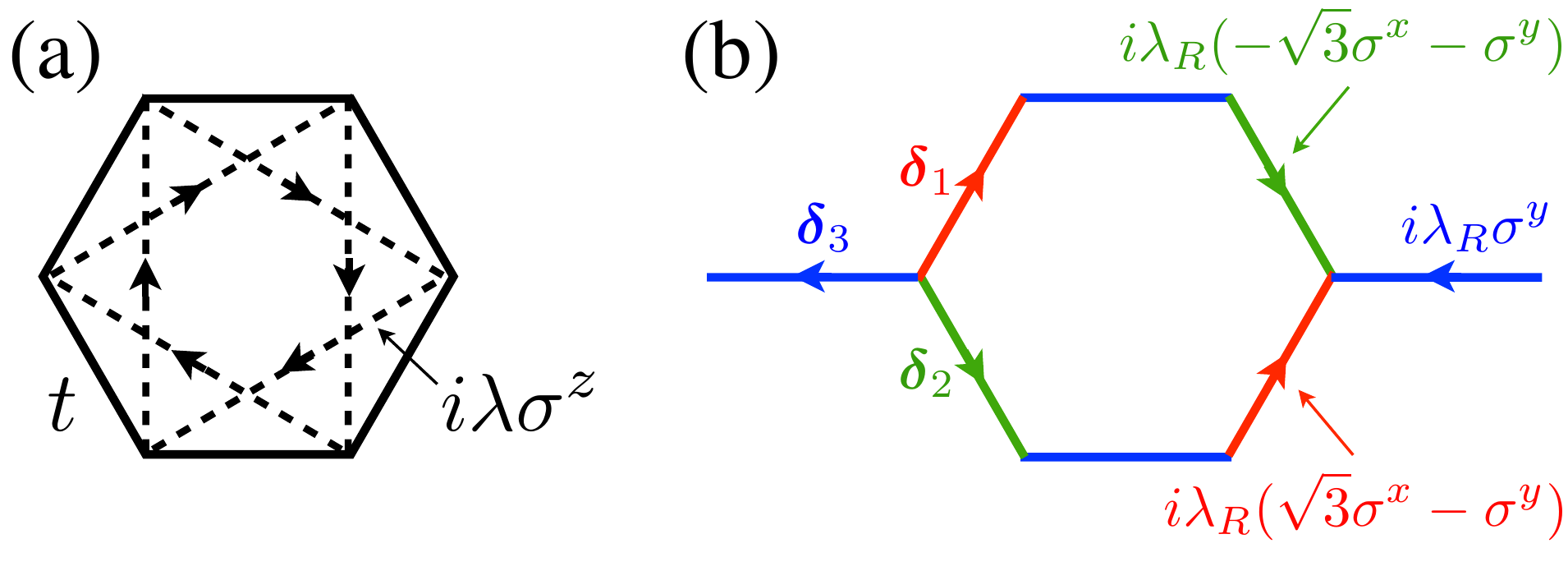}
    \caption{(Color online). (a)  Illustration of the hopping term $\propto t$ and the intrinsic SO term $\propto i\lambda\sigma^z$. (b) Illustration of the nearest-neighbor vectors $\bs{\delta}_i$ ($i=1,2,3$) and of the Rashba SO term $\propto i\lambda_R$ with different spin-dependences in different hopping directions $\bs{\delta}_i$.}
    \label{fig:soc}
\end{figure}

%
\section{Model and Methodology}

\subsection{Kane-Mele Hubbard model with Rashba spin-orbit coupling}

The Kane-Mele-Hubbard model is governed by the Hamiltonian
\begin{equation}\label{ham}
\begin{split}
\mathcal{H} =& -t \sum_{\langle ij \rangle\,\sigma} c_{i\sigma}^\dag c_{j\sigma}^\pd +i\lambda \sum_{\langle\!\langle ij \rangle\!\rangle\,\alpha\beta} c_{i\alpha}^\dag \nu_{ij} \sigma^z_{\alpha\beta} c_{j\beta}^\pd \\[5pt]
& + i \lambda_R \sum_{\langle ij \rangle\,\alpha\beta} c_{i\alpha}^\dag (\bs{\sigma}_{\alpha\beta} \times \bs{d})_z \,c_{j\beta}^\pd +U\sum_i n_{i\up}n_{i\dw}\ .
\end{split}
\end{equation}
The operator $c_{i\alpha}$ annihilates a particle with spin $\alpha$ on site $i$, $t$ is the hopping amplitude (which we set  to unity, $t\equiv 1$, throughout the paper), $\lambda$ the intrinsic spin orbit coupling, $\lambda_R$ the amplitude of the Rashba SO coupling,  $U$ parametrizes the local Coulomb (Hubbard) interactions, and $\nu_{ij}=\pm 1$ depending on whether the electron traversing from $i$ to $j$ makes a right (+1) or a left turn (-1) (Fig.~\ref{fig:soc}a). As usual, $\langle ij \rangle$ indicates that $i$ and $j$ are nearest-neighbor sites while $\langle\!\langle ij \rangle\!\rangle$ refers to second-nearest neighbors.
The vector $\bs{d}$ points from site $i$ to site $j$ and corresponds to the nearest-neighbor vectors $\bs{\delta}_i$, ($i=1,2,3$) (Fig.\,\ref{fig:soc}\,(b)); $\sigma^\mu$ ($\mu=x,y,z$) denotes the three Pauli matrices corresponding to spin degree of freedom. 
The explicit spin dependence of the Rashba SO term, $(\bs{\sigma}\times\bs{d})_z$, is visualized in Fig.\,\ref{fig:soc}\,b.
The spin orbit term $\propto\lambda$ breaks the SU(2) symmetry down to U(1), the Rashba term $\propto\lambda_R$ breaks the remaining U(1) spin symmetry down to $\mathbb{Z}_2$.  
It also breaks the spatial inversion symmetry explicitly.
The Rashba spin-orbit term as a part of the original Kane-Mele model has so far generally been neglected in studies of the interacting scenario.
Note that in the original work by Kane and Mele, also a staggered sublattice potential (Semenoff mass) has been discussed which we will not elaborate on further in the following. This term is particularly useful to probe the transition from a topological band insulator phase into a trivial band insulator phase~\cite{haldane88prl2015,kane-05prl146802,kane-05prl226801,cocks-12prl205303,orth-13jpb134004,rachel13arXiv:1310.3159}, but does not yield distinctly new phases, which is the focus of our investigations in the following.

%
%
\subsection{Variational Cluster Approach}

\subsubsection{Method}

The zero temperature variational cluster approach (VCA)\,\cite{potthoff03epjb335} is based on the self-energy functional theory\,\cite{potthoff03epjb429,potthoff05assp135}, which provides an efficient
numerical technique for studying strongly correlated  
systems, especially in the presence of different competing, potentially long-ranged, orders.   
VCA simplifies the lattice problem, as defined in
Eq.~\eqref{ham}, to an exactly solvable problem
defined in a reference system consisting of decoupled finite-size
clusters.
The thermodynamic limit is recovered by reintroducing the inter-cluster hopping to the decoupled cluster  via a non-perturbative
variational scheme based on self-energy functional theory.
The VCA has been successfully applied to many interesting problems,
including the high-T$_{c} $ cuprates~\cite{senechal-05prl156404,balzer-10prb144516} and
correlated topological insulators~\cite{yu-11prl010401}.
In particular, this method is suitable for our current study since the topologically non-trivial properties of the $\mathbb{Z}_2$ topological insulators are appropriately accounted for. By construction, the VCA becomes exact in the limit of $U\to 0$. Hubbard onsite interactions might give rise to competing phases (such as magnetic order) which  can be accurately described by the
VCA grand potential.

\begin{figure}[t]
        \begin{center}
                \includegraphics[width=\linewidth]{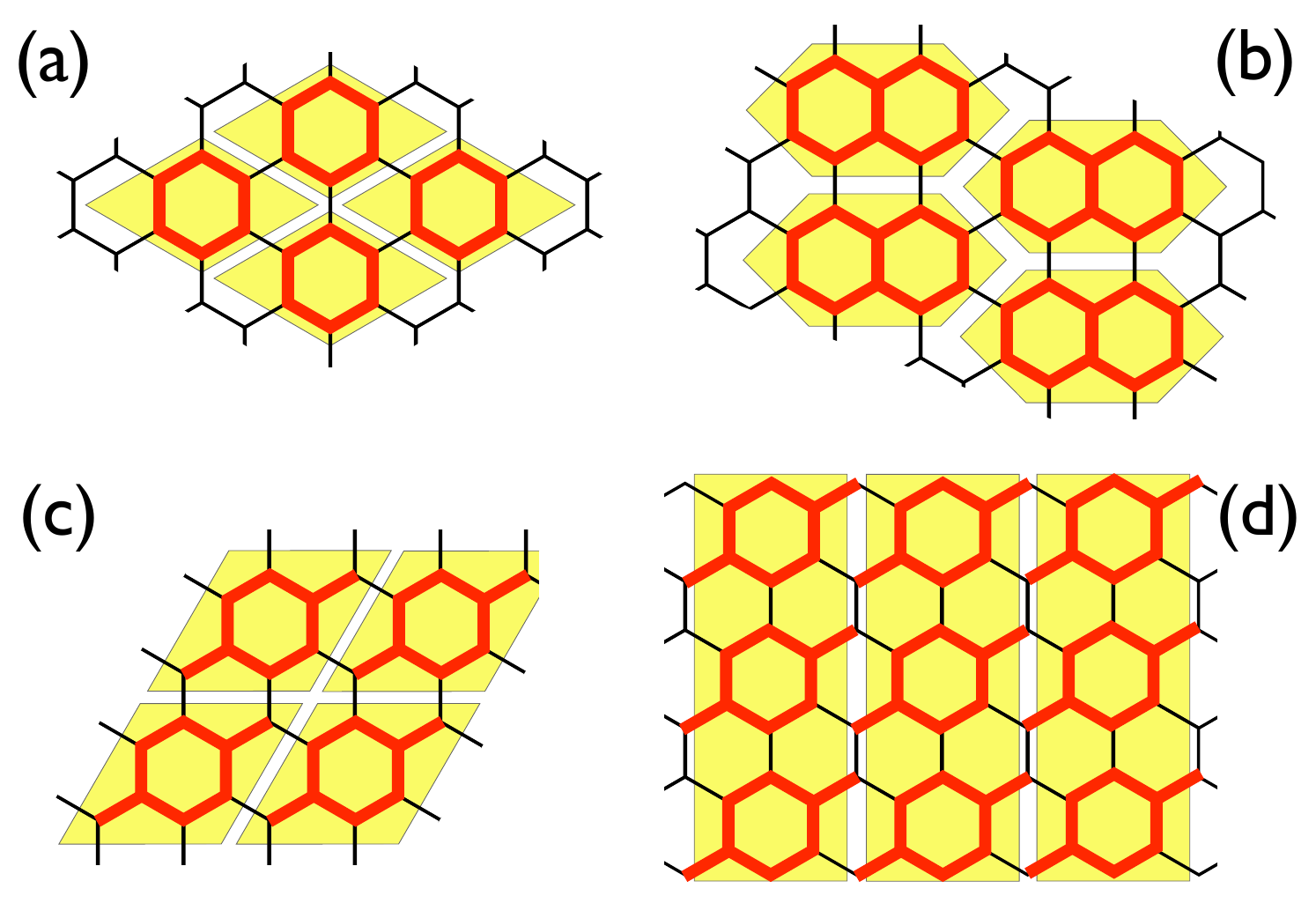}
        \end{center}
    \caption{(Color online). Honeycomb lattice covered with single clusters in VCA: (a) six-site clusters (PBC). (b) ten-site clusters (PBC). (c) eight-site clusters (PBC). (d) Honeycomb ribbon (cylinder) covered with eight-site clusters.}
        \label{fig:clusters}
\end{figure}

In the self-energy functional theory, the grand potential of a system defined by a Hamiltonian
$H=H_0(\mathbf{t})+H_1(\mathbf{U})$ is written as
a functional of the self-energy $\Sigma$: 
\begin{align}
  \Omega[\Sigma]&= F\left[ \Sigma \right]+\text{Tr}\ln\left(
  G^{-1}_0-\Sigma \right)^{-1} \, , 
  \label{eq:grand-potential-functional}
\end{align}
where $F\left[ \Sigma \right]$ is the Legendre transform of the
Luttinger-Ward functional and $G_0=(\omega+\mu-\mathbf{t})^{-1}$ is
the non-interacting Green's function.
It can be shown that the functional $\Omega[\Sigma]$ becomes
stationary at the physical self-energy, \ie $\delta\Omega\left[
  \Sigma_{\rm phys} \right]=0$.\cite{potthoff03epjb335} 
As the Luttinger-Ward functional is universal, it has the
same interaction dependence for systems with any set of $\mathbf{t'}$
as long as the interaction $\mathbf{U}$ remains unchanged. 
Note that the functional $\Omega\left[ \Sigma \right]$ itself is not approximated by any means; we restrict, however, the ``parameter'' space of possible self-energies to the self-energies of the reference system.
Thus, the stationary points are obtained from the self-energy
$\Sigma'=\Sigma\left[ \mathbf{t'} \right]$ of a system defined by the Hamiltonian
$H^{\prime}=H_0(\mathbf{t'})+H_1(\mathbf{U})$, which we label as reference
system. Let us define $V=\mathbf{t}-\mathbf{t}'$. Now we are able to conveniently define the VCA-Green's function,
\begin{equation}
G_{\rm VCA}^{-1} = G'^{-1} - V\ .
\end{equation}
In terms of the reference system, the VCA grand potential is
calculated more conveniently as 
\begin{align}
  \Omega[\Sigma']&= \Omega'+\text{Tr}\ln\left( G^{-1}_0-\Sigma'
  \right)^{-1} - \text{Tr}\ln(G') \, ,
  \label{eq:grand-potential}
\end{align}
with $\Omega'$, $\Sigma'$, and $G'$ denoting the grand potential, the
self-energy and the Green's function of the reference system, respectively.
The reference system is chosen such that it can be treated exactly. Here, we choose an array of decoupled
clusters with open boundary conditions and calculate $\Omega'$, $\Sigma'$, and $G'$ via exact
diagonalization. 
While the correlation beyond the reference system size are included on a mean-field level,
the short-range correlations within the reference
system are fully taken into account in the VCA, resembling related (cluster) DMFT approaches.

%
%
\subsubsection{Cluster size and shape}

Since a spinful Hubbard model involves four basis states for each lattice site, we are generally restricted to rather small clusters with a maximum of ten sites  (Fig.\,\ref{fig:clusters}\,(b)). Furthermore, the choice of the reference system, \ie the cluster
shape and size, {is constrained by the requirement that the honeycomb lattice needs to be fully covered, either using periodic boundary conditions (PBCs)--as realized on a torus--or cylindrical boundary conditions.} We consider six-, eight-, and ten-site clusters in the case of PBCs  and eight-site clusters for cylindrical boundary conditions with zig-zag edges (Fig.\,\ref{fig:clusters}). (Note that the six- and ten-site clusters could also be used for ribbons (cylinders)  with armchair edges which is not further considered here, see also Ref.\,\onlinecite{wu-12prb205102}.) While one generally expects to obtain more accurate results with a larger cluster, the effect of the lattice partitioning, \ie the cluster dependence, is rather strong. {\it We therefore extract our physical results from the joint consideration of all cluster sizes reachable by VCA, which is indispensable to obtain physically meaningful results from finite cluster approaches in general.}

In the topological insulator phase we explore the edge states connecting the valence and conduction bands of the system. These edge states typically penetrate a few unit cells into the bulk.
If the ribbon height (\ie the distance between upper and lower edge) does not exceed a few unit cells it might happen that the penetrating edge states from the upper and lower edge couple to each other and gap out. 
To avoid this, we have to make sure that the ribbon height is sufficiently large; we build a supercluster which consists of $n$ normal clusters (as described above) and stack them on top of each other
 as illustrated in Fig.\ref{fig:clusters}\,(d). The supercluster corresponds to the unit cell of the effectively one-dimensional superlattice and is defined by the tridiagonal matrix
\begin{equation}
  G'^{-1}=\begin{pmatrix}
    G'^{-1}_1& t_{1,2}    \\[5pt]
    t_{2,1} &~G'^{-1}_2~& t_{2,3}  \\[5pt]
     & t_{3,2}& ~G'^{-1}_3~& t_{3,4}  \\[5pt]
     &&\ddots&\ddots&\ddots \\[5pt]
    &&&t_{n-1,n-2}&~G'^{-1}_{n-1}~ &t_{n-1,n} \\[5pt]
   &&&&t_{n,n-1}&G'^{-1}_{n} \end{pmatrix}
  \label{eq:supercluster}
\end{equation}
where $G'$ is the Green's function of the supercluster with the dimension $2L_c \times n$, $G'_{i}$ are the cluster Green's functions and $t_{i,i+1}$ is the hopping matrix connecting the two cluster Green's functions $G'_i$ and $G'_{i+1}$; $L_c$ is the number of cluster sites.
To separate edge states from the upper and lower edge we stack at least eight clusters to form a supercluster from which we compute the single-particle spectral function (displaying the edge states). The single-particle spectral function $A(k,\omega)$ is defined as in the standard case of PBCs via
\begin{equation}\label{def-Akw}
A(k,\omega) = -\frac{1}{\pi} {\rm Im}\Big\{ G_{\rm VCA}(k,\omega) \Big\}\ ,
\end{equation}
where the VCA-Green's function depends on the momentum $k$ retained by the circumferential direction of the cylinder.

%
%
\subsubsection{Symmetry breaking Weiss fields}
In quantum cluster approaches (and dynamical mean-field theory) manifestations of spontaneous symmetry breaking for finite size clusters is resolved by introducing artificial mean-field like Weiss fields of the form
\begin{equation}\label{xaf-weiss}
H_{X-{\rm AF}} = h^x \sum_{i\,\alpha\beta} \left( a_{i\alpha}^\dag \sigma^x_{\alpha\beta} a_{i\beta}^\pd - b_{i\alpha}^\dag \sigma^x_{\alpha\beta} b_{i\beta}^\pd \right)\ ,
\end{equation}
where the operator $a_i$ ($b_i$) acts on sublattice $A$ ($B$).
Eq.\,\eqref{xaf-weiss} is the simplest example of an antiferromagnetic Weiss field with N\'eel order in $x$-direction (in-plane). Given an external Weiss field for a certain order parameter, a stable magnetic solution is characterized by a stationary point in the grand potential at a finite field strength. Furthermore, in order to represent the physical ground state, such a stationary point needs to have a lower energy than the zero-field solution. In principle, similar to a mean-field treatment, this procedure needs to be repeated for all possible configurations of Weiss fields. The order parameter can then be determined from the magnetic solution with the lowest energy. The cluster decomposition of the lattice, however, restricts the possible choices of Weiss fields to those which are compatible with the cluster size and shape, \ie a Weiss field needs to have the same periodicity as the array of clusters. Typically, for a given cluster only a few types of magnetic order may be investigated. For example, a N\'eel pattern cannot be implemented on a three-site cluster. Likewise, incommensurate spiral order is incompatible with any finite cluster.

%
%
\subsubsection{Variation of single-particle parameters}
The variational procedure of VCA works such that the amplitudes of every single-particle term as well as the chemical potential $\delta\mu$ need to be varied. It is well established, however, that for practical purposes the variation of $\delta\mu$ is often sufficient and the additional variation of, say, the hopping $\delta t$ does not lead to a new stationary point. For the KMH model, in principle we have to vary not only the chemical potential, but also hopping, spin orbit coupling, and Rashba term independently. In the Appendices A and B, we show exemplarily the difference between (i) variation of $\delta\mu$, (ii) variation of $\delta\mu$ and $\delta t$, (iii) variation of  $\delta\mu$, $\delta t$, and $\delta \lambda$, as well as (iv) variation of additional antiferromagnetic Weiss fields. Essentially we find that variation of $\delta t$ has a significant effect on the phase diagrams incl.\ magnetic phase transitions. Additional variation of $\delta\lambda$ or $\delta\lambda_R$, respectively, does not seem to influence the variational procedure. Still, performing VCA on the honeycomb lattice with variation of $\delta\mu$ only might lead to numerical artifacts and should be avoided. Further details are illustrated in the Appendices A and B.

%
%
\section{Kane-Mele-Hubbard model without Rashba SO Coupling $\bs{(\lambda_R=0)}$}

%
%

\subsection{Topological insulator}

\subsubsection{$\mathbb{Z}_2$ invariant}

In the presence of inversion symmetry the topological invariant can be conveniently calculated probing bulk properties only, which is even applicable in the interacting case. Particularly, within VCA this can be achieved for any cluster size.

Expressing topological invariants in terms of single particle Green's functions was pioneered by Volovik~\cite{volovik03}; more recently, Gurarie\,\cite{gurarie-11prb085426} conveniently reformulated Volovik's invariant for the field of topological insulators. Recently, Wang \ea\,\cite{wang-10prl256803,wang-12prb165126} derived simplified expressions for the inversion-symmetric Hamiltonians. The $\mathbb{Z}_2$ topological invariant relevant for topological insulators is computed from the full interacting Green's function  through a  Wess-Zumino-Witten term\,\cite{wang-10prl256803}, motivated from the concept of dimensional reduction in topological field theory~\cite{qi-11rmp1057,qi-08prb195424}. 

\begin{figure}[t]
\begin{center}
\includegraphics[width=.35\textwidth]{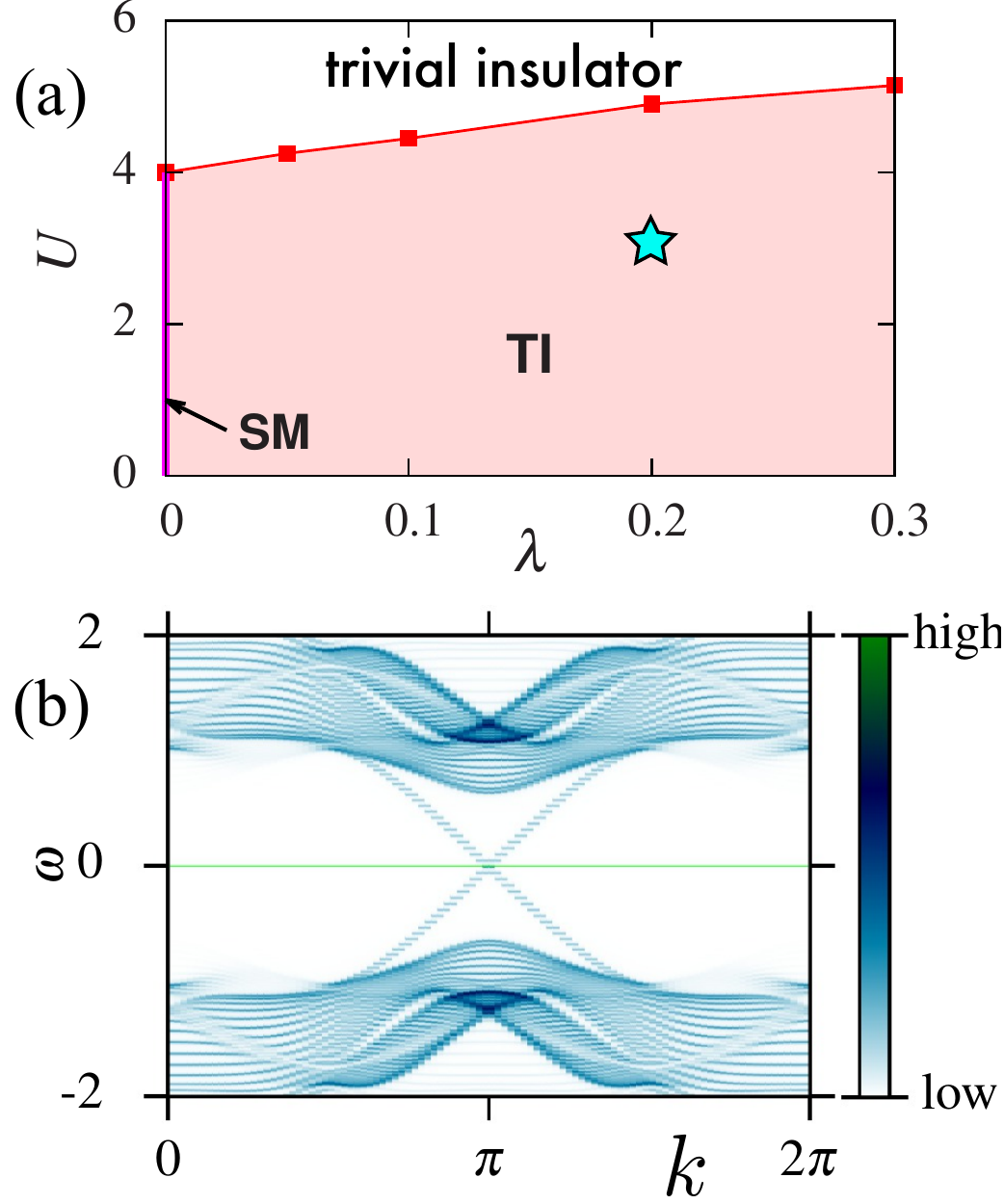}     \end{center}
\caption{(Color online). (a) Phase boundary in $U$--$\lambda$--plane between topological insulator and trivial band-insulator (``non-magnetic'' solution) obtained by a periodic eight-site cluster computation of the  $\mathbb{Z}_2$ invariant. 
(b) Edge spectrum in the TI phase obtained for cylindrical geometry; parameters ($\lambda=0.2$, $U=3$, $\lambda_R=0$) correspond to the light-blue star in the phase diagram (a). (a) and (b) are complementary approaches to detect the topological insulating phase.}
\label{fig:kmh-z2invariant}
\end{figure} 

In the presence of  inversion symmetry (\ie when $\lambda_R\equiv 0$ and antiferromagnetic order is absent), we follow Wang \ea to compute the topological invariant formula\,\cite{wang-12prb165126} via the parity eigenvalues of the Green's function obtained within VCA at the time-reversal invariant momenta (TRIM) $\bs{\Gamma}_{i}$ and zero energy. 
The Green's function is a $N\times N$ matrix with $N=2L_c$, where $L_c$ is the number of sites per cluster. Both $G$ and $G^{-1}$ can be diagonalized, yielding
\begin{equation}
  G(i\omega,\bs{k})^{-1} \ket{\alpha(i\omega,\bs{k})}  = \mu_\alpha(i\omega,\bs{k}) \ket{\alpha  (i\omega,\bs{k})}\ ,
  \label{eq:eigen-green}
\end{equation}
with $\mu_\alpha \in \mathbb{C}$. The Green's function matrix $G(i\omega,\bs{k})$ has the same eigenvectors $\ket{\alpha(i\omega,\bs{k})}$ but the inverse eigenvalues $\mu^{-1}_\alpha(i\omega,\bs{k})$. The states at the TRIMs, $\ket{\alpha  (i\omega,\bs{\Gamma}_i)}$, are simultaneous eigenstates of $G$ and $P$  and satisfy\,\cite{wang-12prb165126},
\begin{equation}
  P \ket{\alpha  (i\omega,\bs{\Gamma}_i)} = \eta_\alpha \ket{\alpha  (i\omega,\bs{\Gamma}_i)}\ .
  \label{eq:eigen-parity}
\end{equation}
Since $\mu_\alpha(0,\bs{\Gamma}_i)$ is real, one can distinguish between positive ($\mu_\alpha(0,\bs{\Gamma}_i)>0$) and negative ($\mu_\alpha(0,\bs{\Gamma}_i)<0$) eigenvalues, denoted as R-zeros and L-zeros, respectively. This allows to define the topological invariant $\Delta$ via
\begin{equation}
  (-1)^{\Delta}=\prod_{\rm R-zero} \eta_{\alpha}^{1/2}=\pm1\ .
  \label{eq:ti}
\end{equation}
In Fig.\,\ref{fig:kmh-z2invariant}\,(a) we show the $U$--$\lambda$ plot of this invariant. 
Note again that $\Delta$ cannot be calculated when an antiferromagnetic Weiss field is present due to breaking of inversion symmetry.
As a consequence, in VCA we independently investigate the magnetically ordered regime. 
The onset of a finite magnetization likewise sets the boundary for which the topological character of the insulating state vanishes.

\subsubsection{Edge states}

As an alternative to a bulk measurement of the topological invariant, the topological insulator phase can also be identified by detecting the helical edge states which are a hallmark of $\mathbb{Z}_2$ topological insulators considered here. This is accomplished by solving the Hamiltonian \eqref{ham} on a cylindric geometry as explained in the previous section. This method is reliable and is also applicable when the computation of the topological bulk invariant is too complicated, such as for finite Rashba SO coupling addressed later. In Fig.\,\ref{fig:kmh-z2invariant}\,(b) the single particle spectral function $A(k,\omega)$ defined for a ribbon geometry is shown ($\lambda=0.2$, $\lambda_R=0$, $U=4$).
In the effectively one-dimensional Brillouin zone, one clearly sees a band gap between upper and lower bands, which are connected by helical edge states crossing at the TRIM $k=\pi$.

\subsection{XY Antiferromagnet}

For $\lambda\to 0$ the Hamiltonian \eqref{ham} becomes invariant under SU(2) spin rotations
and the antiferromagnetic N\'eel order is isotropic. 
Finite SO coupling $\lambda\not= 0$ drives the system into an easy-plane antiferromagnet with an ordering vector in the $x$-$y$ lattice plane\,\cite{rachel-10prb075106}, which has been confirmed by QMC\,\cite{hohenadler-11prl100403,dong-11prb205121}, VCA\,\cite{yu-11prl010401}, and pseudofermion functional RG\,\cite{reuther-12prb155127}.
In order to compute the magnetic phase diagram within VCA, we apply antiferromagnetic Weiss-fields in $x$ and $z$-direction for various values of $\lambda$.

\begin{figure}[t]
\begin{center}
\includegraphics[width=.45\textwidth]{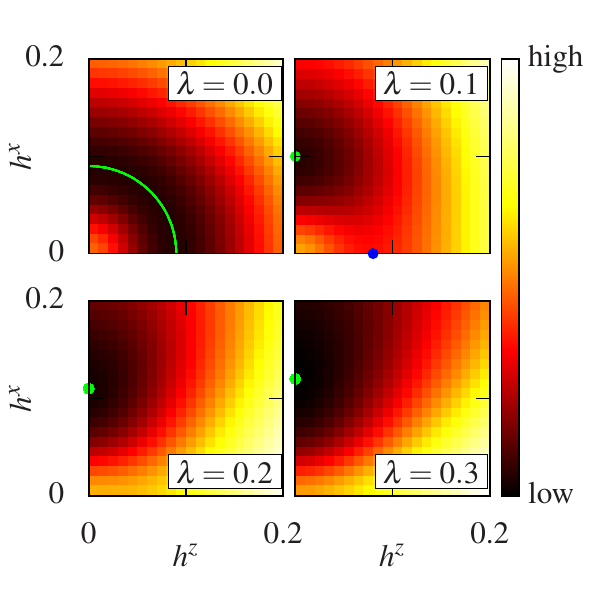}
\end{center}
\caption{(Color online). Heat map of the grand potential $\Omega(h^{x},h^{z})$ as a function of  antiferromagnetic Weiss fields $h^{x}$ and $h^{z}$ for various values of $\lambda$.
All plots haven been obtained for the six-site cluster and $U=6$.
Global minima of $\Omega$ are indicated by green points (lines).
For $\lambda=0.1$ we find a second stationary point (blue point) which is a saddle point at finite $h^z\ne 0$ with higher energy.}
\label{fig:af-xz-so}
\end{figure}

For $\lambda=0$ we find a circle of degenerate minima in the $h^x$-$h^z$-plane, indicating isotropic magnetic order.
For finite $\lambda>0$, this degeneracy is lifted and magnetic order in $x$-direction is energetically preferred. For small $\lambda=0.1$ there is an additional stable solution (a saddle point in $\Omega$ indicated by the blue point in Fig.\,\ref{fig:af-xz-so} right top panel) corresponding to a magnetization in $z$-direction. This solution, however, is not a global minimum in $\Omega$ and the system is still an easy-plane antiferromagnet. For larger $\lambda$, this meta-stable solution disappears. 
In total, the VCA confirms the established results about magnetic order in the KMH.

\subsection{Phase diagram}

\begin{figure}[t]
    \begin{center}
        \includegraphics[width=.4\textwidth]{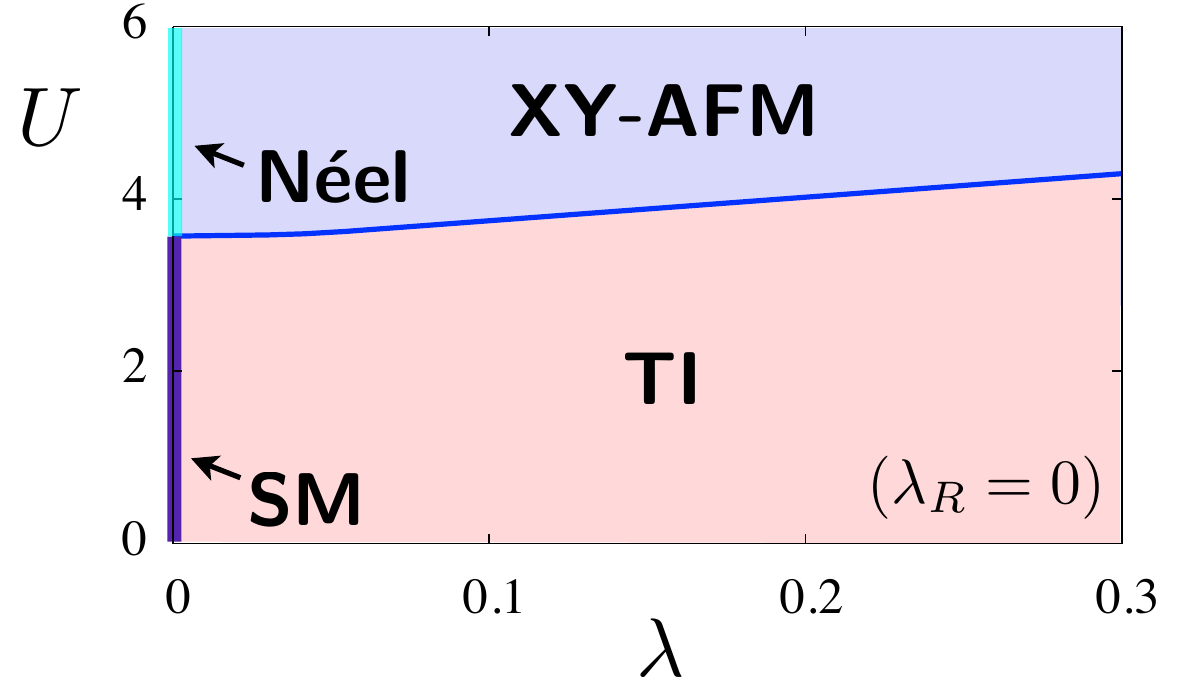}
    \end{center}
    \caption{(Color online). Schematic phase diagram of the Kane-Mele-Hubbard model ($\lambda_R=0$) as obtained from VCA.}
    \label{fig:kmh-schematic}
\end{figure}

As the final result, the interacting $U$--$\lambda$ phase diagram exhibits a semi-metal for $\lambda=0$ which is detected via a linear density of states near the Fermi-level.
It transcends into a topological insulator phase for finite $\lambda$ up to moderate interaction strengths. For stronger interactions, the system 
acquires XY antiferromagnetic order. Obtaining a phase diagram such as Fig.\,\ref{fig:kmh-schematic} via a quantum cluster approach is challenging:
(i) stabilizing semi-metals within real-space quantum cluster methods is rather involved; in particular the six-site cluster may suffer from artifacts of the lattice partitioning.
(ii) clusters which do not have the shape of closed honeycomb rings underestimate the critical interaction strength $U_c$ associated with the onset of magnetization. (iii) exclusive variation of the chemical potential might lead to an erroneous non-magnetic insulator phase up to small intrinsic spin orbit coupling\,\cite{yu-11prl010401}. In our analysis where we also varied the hopping in order to minimize the grand potential we could not find this non-magnetic insulator phase.
Note that this erroneous non-magnetic insulator phase was linked to a proposed quantum spin liquid phase. Recently, it was shown using large-scale QMC calculations that there is no such spin liquid on the honeycomb lattice\,\cite{sorella-12sr992,assaad-13prx031010} being in perfect agreement with our analysis.
(For an extensive discussion and details about (i) -- (iii) we refer the interested reader to Appendix A.)
The analysis done so far shows that a careful multi-size cluster analysis has to be employed in order to determine an artefact-free physical phase diagram. This equips us for our subsequent investigations of the KMH model in the presence of Rashba SO coupling studied in the next section.

%
%

\section{Kane-Mele-Hubbard model including Rashba SO coupling $\bs{(\lambda_R>0)}$}

\begin{figure*}[t]
\begin{center}
\includegraphics[width=1.0\textwidth]{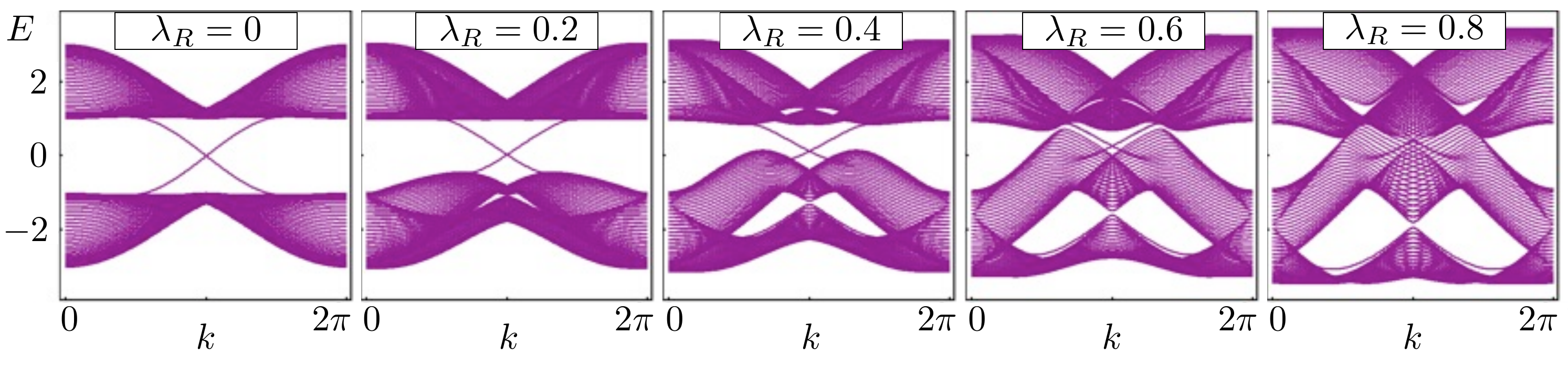}
\caption{(Color online). Single particle spectra on a cylinder geometry for $U=0$, $\lambda=0.2$, and different values of $\lambda_R$. From left to right: $\lambda_R=0$, $0.2$, $0.4$, $0.6$. and $0.8$. The spectra interpolate from a topological insulating phase ($\lambda_R=0$, $0.2$, and $0.4$) to a metallic phase ($\lambda_R=0.8$).
In between, for $\lambda_R=0.6$ we find an additional weak topological semiconductor phase (see also Fig.\,\ref{fig:ts-phase}).}
\label{fig:akw-U=0}
\end{center}
\end{figure*} 

\begin{figure}[t]
\begin{center}
\includegraphics[width=\linewidth]{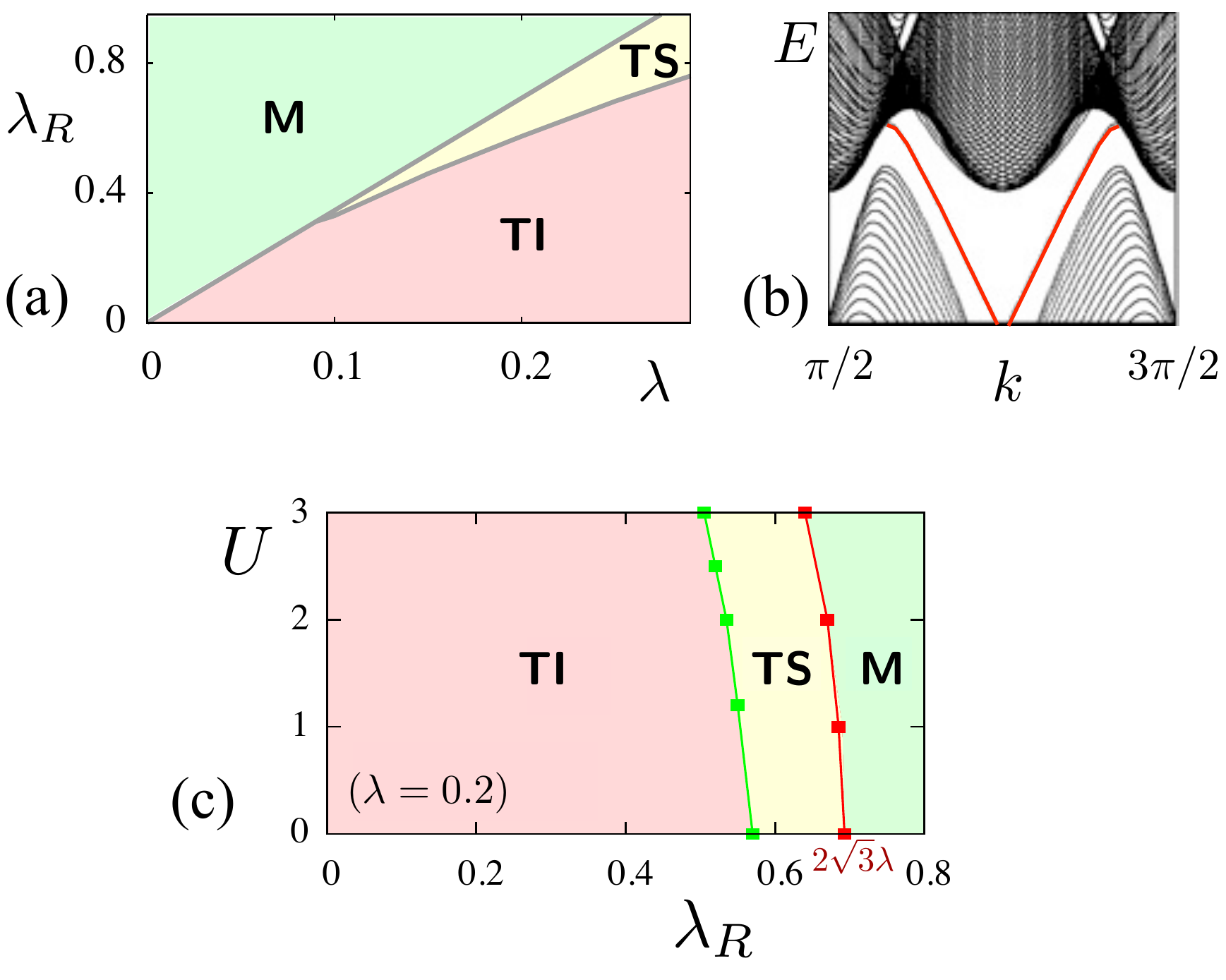}
\caption{(Color online).  (a) $\lambda_R$--$\lambda$ phase diagram for the non-interacting Kane-Mele model displaying the TI, metal (M), and topological semi-conductor (TS) phase. (b) Zoom into the edge spectrum for $\lambda=0.2$, $\lambda_R=0.6$, $U=0$ shown in Fig.\,\ref{fig:akw-U=0}. 
(c) $U$--$\lambda_R$ phase diagram for $\lambda=0.2$ in the non-magnetic regime: the weak TS phase persists in the presence of interactions.}
\label{fig:ts-phase}
\end{center}
\end{figure}

In their seminal papers, Kane and Mele showed that the topological insulator phase persists until $\lambda_R = 2\sqrt{3}\lambda$ where the gap closes and the system enters a metallic phase\,\cite{kane-05prl146802,kane-05prl226801}. They computed the $\mathbb{Z}_2$ invariant to explore the corresponding phase diagram. 
In their work, they considered rather small values of SO coupling such as $\lambda = 0.03$ or $0.06$, and in general $\lambda\ll t$. 
For a description of graphene, which was the original intention of this work, such small SO coupling seemed to be realistic.
However, with regard to the many different candidate systems potentially realizing the quantum spin Hall effect in a honeycomb lattice compound which have been proposed in the meantime, it is justified to consider larger spin orbit coupling such as $\lambda=0.2$. It turns out, that for sufficiently large $\lambda\geq 0.1$ and $\lambda_R$ close to the predicted phase transition at $\lambda_R=2\sqrt{3}\lambda$, the system is not gapped anymore. The Rashba SO coupling bends the bands such that there is no full gap. On the other hand, there is always a direct gap for each wave vector $k$, \ie the conductance and valence bands do neither touch nor cross each other -- this is the reason why the topological invariant (computed for $U=0$) labels this region as a topological insulator. In fact, in this ``metallic'' region the edge states are well-defined and clearly visible (see the second-right panel in Fig.\,\ref{fig:akw-U=0} and Fig.\,\ref{fig:ts-phase}\,(b)). At each momentum $k$ the system has a gap, but globally the system is gapless. Therefore we call this region a 
weak topological semiconductor phase where ``semiconductor'' refers to a {\it direct gap-only insulating phase}. In the presence of disorder individual $k$ values cannot be distinguished anymore leading to the attribute {\it weak}. Still this phase is stable in the presence of interactions as we will explicate below.

%
%
\subsection{Weak to intermediate interactions}

\begin{figure}[t]
    \centering
        \includegraphics[width=.37\textwidth]{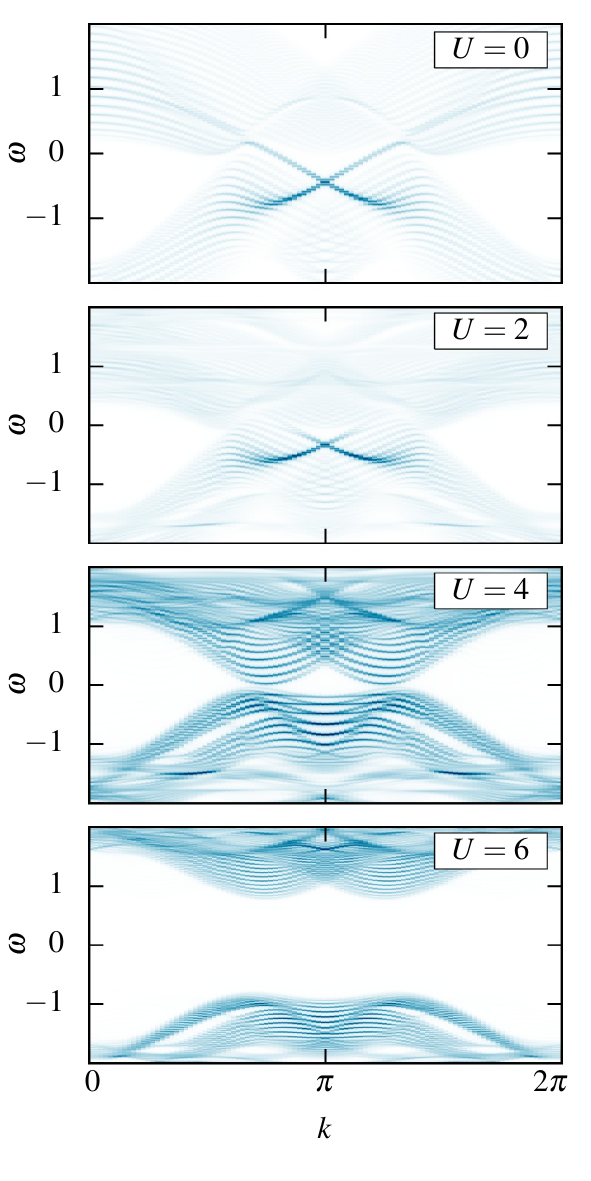}
     \caption{(Color online). Spectral function $A(k,\omega)$ on cylindrical geometry (as defined in Eq.\,\eqref{def-Akw}) for $\lambda=0.2$, $\lambda_{R}=0.6$, and various values of $U$. For better illustration, only the weights of the outermost sites on the cylinder are taken into account. From top to bottom: $U=0$, $2$, $4$, and $6$. For $U=0$ and $U=2$ we find the weak TS phase, for $U=4$ and $U=6$ a magnetically ordered insulating phase.}
    \label{fig:spec-multi-km01-ra06-u}
\end{figure}

For $\lambda<0.1$, we only find  TI and metallic phases at $U=0$, which persist for moderate interaction strength.
Fixing $\lambda=0.2$ we find three different phases at $U=0$: TI, weak topological semiconductor (TS) phase, and metal (see Fig.\,\ref{fig:ts-phase}\,(a,b)). 
The TS phase is stable with respect to interactions, see Fig.\,\ref{fig:ts-phase}\,(c). To gain further insight, we compute single-particle spectral functions on cylindrical geometry (using the eight-site cluster) to determine the edge state spectrum (see Fig.\,\ref{fig:spec-multi-km01-ra06-u}). For $\lambda=0.2$ and $\lambda_R=0.6$, the TS phase is stable up to moderate values of $U$. 
At around $U=4$ the system enters a magnetically ordered phase. Upon further increasing $U$ the bulk gap increases rapidly; however, no edge states connect the valence and conductance bands anymore, indicating the trivial topology of the magnetic phase. 

We perform an additional test to verify that the two modes crossing at $k=\pi$ in Fig.\,\ref{fig:spec-multi-km01-ra06-u} ($U=0$ and $U=2$) are indeed 
edge states:
we repeat the computation of the single particle spectral function $A(k,\omega)$ on a cylindrical geometry but with additional links connecting the two edges of the cylinder. These additional links are chosen such that they are compatible with the band structure of the KMH model. As such, moving from a cylindric to a toroidal geometry, the bulk spectra should be unchanged with the only difference that the edges have disappeared, which is exactly what we find.


%
%
\subsection{Strong interactions and magnetic order}

For finite $\lambda>0$ and $\lambda_R=0$, the magnetic region of the phase diagram is an XY antiferromagnet as discussed above. Treating the Rashba term as a small perturbation leaves the magnetic phase unchanged. Thus we expect  an XY-AFM in the weak-$\lambda_R$ region.
\begin{figure}[t!]
    \begin{center}
        \includegraphics[width=.35\textwidth]{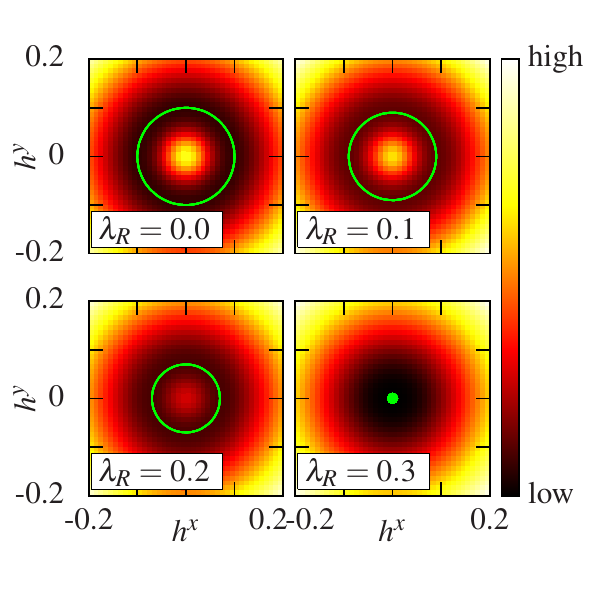}
    \end{center}
    \caption{(Color online). Heat map of the grand potential as a function of antiferromagnetic Weiss fields $\Omega(h^x,h^y)$. On the six-site ring-shaped cluster we find easy-plane AFM order for $\lambda_{R}<0.3$ (at $\lambda=0.1$ and $U=6$). For larger Rashba coupling we do not find any saddle points at finite Weiss fields. }
    \label{fig:af-xy-so-km01-ra00-03}
\end{figure} 
First, we use the six-site cluster and compute the grand potential $\Omega$ as a function of $h^x$ and $h^y$. As expected we find the XY-AFM.  $\Omega$ as a function of $h^x$ and $h^y$ shows a perfect circle at finite Weiss fields $h^{x/y}$ (Fig.\,\ref{fig:af-xy-so-km01-ra00-03}).

For the six-site cluster, the saddle point associated with the XY-AFM phase is found at decreasing Weiss fields $h^{x/y}$ when we increase the Rashba coupling. For $\lambda_R=0.3$ (at fixed $\lambda=0.1$), we do not find any magnetic solution anymore (see lower panels in  Fig.\,\ref{fig:af-xy-so-km01-ra00-03}). This implies that there is  either a true non-magnetic insulator phase or there is a magnetically ordered phase which cannot be detected within VCA. For instance, this is the case for incommensurate spiral order, where the Weiss field is incompatible with the cluster partitioning. A spiral phase is likely to occur since the spin Hamiltonian (\ie the Hamiltonian obtained in the strong coupling limit $U\to\infty$ of Eq.\,\eqref{ham}) contains terms of Dzyaloshinskii-Moriya type\,\cite{reuther-12prb155127}.
Recently, spiral order was also found in a Kane-Mele type model\,\cite{shitade-09prl256403} with multi-directional SO coupling in the presence of strong interactions\,\cite{reuther-12prb155127,kargarian-12prb205124,liu-13arXiv:1307.4597}.

In principle, we cannot rule out the existence of the non-magnetic insulator phase for large $U$ and large Rashba spin orbit coupling. The existence of such a phase would be exciting, in particular, since it could be related to a recently proposed fractionalized quantum spin-Hall phase (dubbed QSH$^\star$)\,\cite{ruegg-12prl046401}.

\subsection{Phase diagram}

As the final result of this section and this paper, the $U$-$\lambda_R$ phase diagram contains, for moderate Rashba SO coupling $\lambda_R$, a TI phase (weak interactions) and an XY-AFM phase (strong interactions). Stronger Rashba SO coupling drives the TI into a metallic phase. If the intrinsic SO coupling $\lambda$ is sufficiently large ($\lambda \geq 0.1$) an additional weak topological semiconductor phase emerges between the TI and the metallic phase. In the strong-interaction regime, we do not find a magnetic solution whose unit cell would be consistent with the available cluster sizes in VCA, a regime which is hence likely to host incommensurate spiral magnetic order. All these findings cumulate in the schematic phase diagram Fig.\,\ref{fig:schematic-phasedia}. 

\section{Conclusions}
We have investigated the effect of Rashba spin orbit coupling in the  Kane-Mele-Hubbard model as a prototypical correlated topological insulator. We have applied the variational cluster approach and determined the phase diagram via the computation of local density of states, magnetization, single particle spectral function, and edge states to detect the topological character. The topological insulating phase persists in the presence of Rashba spin-orbit coupling and interactions. Furthermore, in the strong coupling regime, the Rashba term induces magnetic frustration which leads to incommensurability effects in the magnetic fluctuation profile and is conjectured to predominantly give rise to spiral magnetic phases. Rashba spin orbit coupling also gives rise to peculiar metallic phases. We find a weak topological semiconductor phase, for a wide range of Hubbard interaction strengths as well as intrinsic and Rashba spin orbit couplings. 
It will be exciting to investigate some of these effects in future experiments which exhibit the Rashba term due to external fields or intrinsic environmental effects.

\begin{acknowledgements}
The authors acknowledge discussions with Karyn Le Hur, Martin Hohenadler, Fakher F.\ Assaad, Andreas R\"uegg, Motohiko Ezawa, Tobias Meng, Michael Sing, J\"org Sch\"afer, and Matthias Vojta.
We thank the LRZ Munich and ZIH Dresden for generous allocation of CPU time.
ML is supported by the DFG through FOR 1162. 
JR acknowledges support by the Deutsche Akademie der Naturforscher Leopoldina through grant LPDS 2011-14.
RT is supported by the ERC starting grant TOPOLECTRICS of the European Research Council (ERC-StG-2013-336012).
SR is supported by the DFG through FOR 960, the DFG priority program SPP 1666 ``Topological Insulators'', and by the Helmholtz association through VI-521. 
We thank the Center for Information Services and High Performance Computing (ZIH) at TU Dresden for generous allocations of computer time.
\end{acknowledgements}

%
%
\appendix

\section{Cluster analysis of the KMH model ($\bs{\lambda_R=0}$)}

\begin{figure*}[t]
\centering
\includegraphics[scale=0.55]{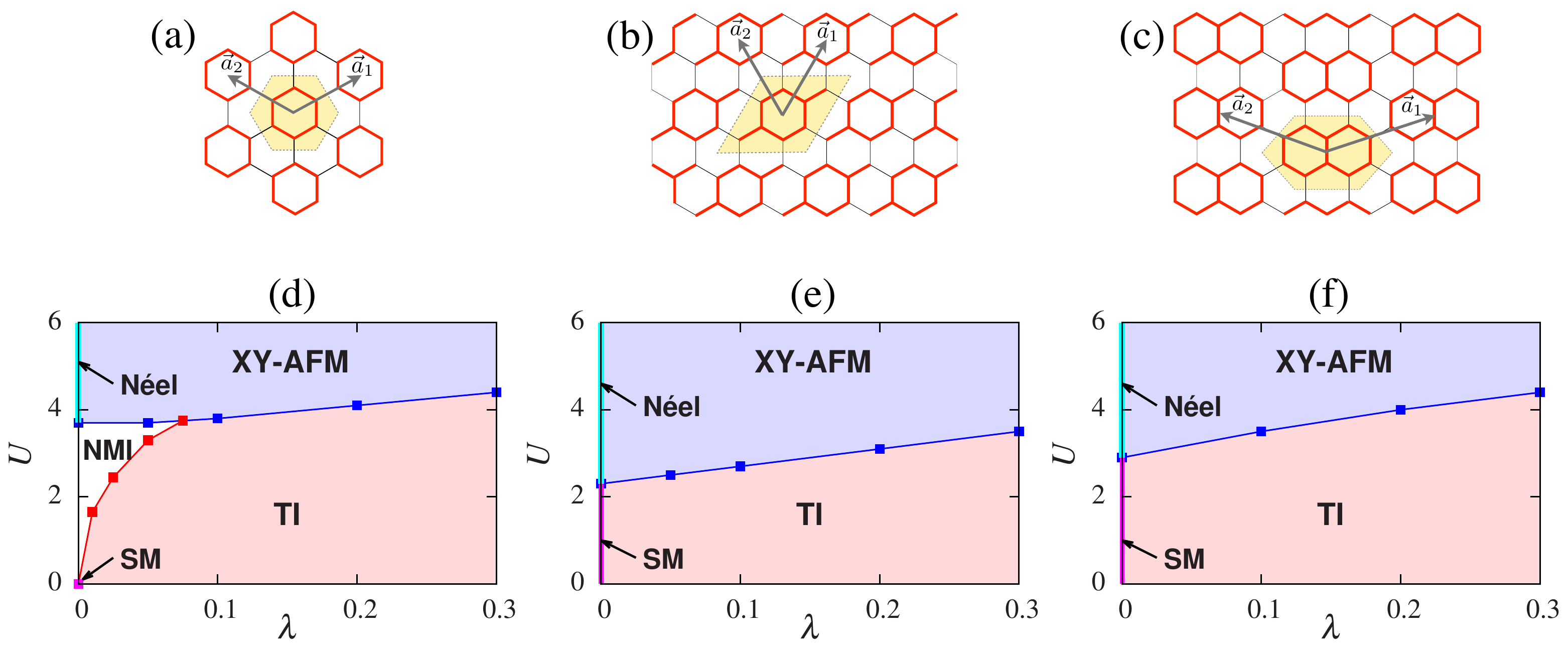}
\caption{(Color online). (a-c) Coupled cluster tight-binding scenarios. Red thick links are associated with $\tilde t$ and black thin lines with $t$. The second-neighbor spin-orbit links are treated analogously but are omitted for clarity of this figure.  (a) Six-site plaquette anisotropic honeycomb lattice 
(b) Eight-site lattice and (c) ten-site lattice. (d-f) Phase diagram of the Kane-Mele-Hubbard model for different cluster sizes.
Note that in the limit $\lambda=0$ the system displays a magnetic N\'eel phase and a semimetal phase for all cluster sizes. (d) Six-site cluster. We find a non-magnetic insulator (NMI), easy plane antiferromagnetic insulator (XY-AFM), and topological insulator (TI). The semi-metal (SM) only exists for $U=0$. The cyan line indicates the onset of magnetic order ($U_{c}=3.8$ for $\lambda=0$). (e) Eight-site cluster. We find SM, TI, and XY-AFM phases. The SM is realized up to $U_c=2.4$ where we observe the onset of magnetization. (f) Ten-site cluster. We find SM, TI, and XY-AFM phases. The SM is realized up to $U_c=2.9$.}
\label{fig:kmh-phasedias}
\end{figure*}

\subsection{Semi-metallic phase for $\bs{\lambda=0}$}

The semi-metal phase of the honeycomb lattice is more sensitive to the lattice partitioning as compared to other phases and lattices. As we will discuss in the following, cluster size and shape influence the results. A six-site cluster (having the shape of a single hexagon, see Fig.\,\ref{fig:clusters}\,(a)) immediately opens a single-particle gap for $U>0$. In contrast, an eight-site cluster (a hexagon with two additional legs, see Fig.\,\ref{fig:clusters}\,(c)) provides an extended semi-metallic region before the gap opens at $U_c$.
It is insightful to further analyze the features of VCA for the different cluster sizes. Let us consider the six-site cluster in the following. As mentioned in Sec.\,II.\,A., one solves the small cluster exactly using exact diagonalization (ED). In the absence of any SO coupling, we expect a semi metallic region for $0<U\leq U_c$ where the effect of the interactions just causes renormalization of the Fermi velocity of the system. In case of our small cluster, we expect a renormalization of the hopping parameter $t$ which we call $\tilde t$. In the next step of the VCA, an (infinitely) large lattice is covered by these ED-clusters, and the clusters are coupled by the hoppings of the original non-interacting bandstructure, \ie by $t$. 
Hereby, the intra-cluster hoppings may be varied in order to find a stationary point in the grand potential. 
That is, for finite but not too large values of $U$, we effectively obtain a plaquette-isotropic honeycomb model\,\cite{wu-12prb205102} as shown in Fig.\,\ref{fig:kmh-phasedias}\,(a).
Remarkably, for nearest neighbor hoppings the band gap opens immediately when $\tilde t \not= t$. Indeed, an infinitesimal anisotropy opens an infinitesimal gap\,\cite{wu-12prb205102}. In agreement with this idea, we find that the VCA method using the six-site cluster finds a semi-metal only for $U=0$. For any finite $U$ a non-magnetic insulator phase appears (Fig.\,\ref{fig:kmh-phasedias}\,(d)). 

We also tested the influence of bath sites for the six-site cluster\,\cite{seki-13arXiv:1209.2101}.
For each correlated site we added one bath site (resulting in an effective 12-site cluster computation). We still found instant opening of the single particle gap, although the size of the gap was reduced compared to the results without bath sites (in agreement with Ref.\,\onlinecite{seki-13arXiv:1209.2101}). Variation of the intra-cluster hoppings $t$ seems to have a similar effect as adding bath sites. Variation of the hoppings and adding bath sites simultaneously further decreases the size of the single particle gap; it does not change, however, the qualitative behavior.

\begin{figure}[t]
    \centering
            \includegraphics[width=.47\textwidth]{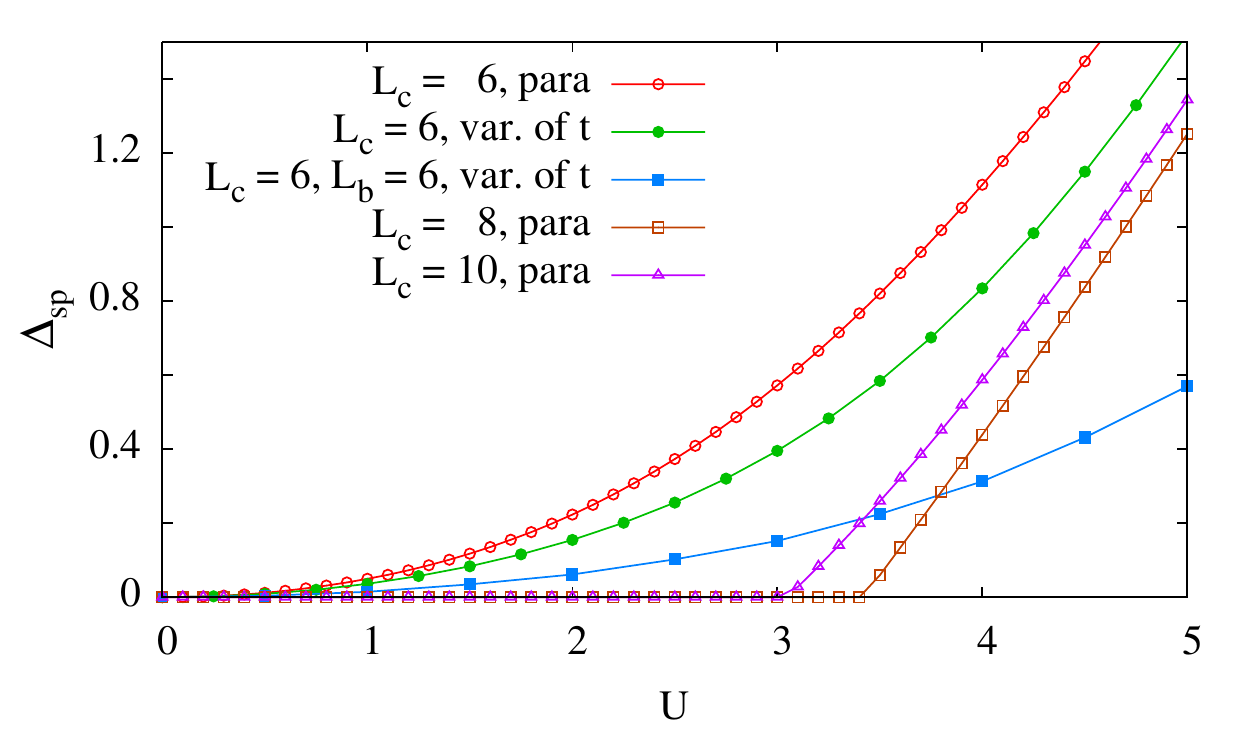}
        \caption{(Color online). Single-particle gap $\Delta_{\rm sp}$ as a function of $U$ ($\lambda=0$) for six-, eight-, and ten-site clusters ($L_c=6,8,10$) with variation of (i) $\delta\mu$ and (ii) $\delta\mu$, $\delta t$. In addition, we show $\Delta_{\rm sp}$ vs.\ $U$ for the six-site cluster with additional bath sites $L_b$ (blue curve). Only the paramagnetic solutions, \ie in the absence of Weiss fields are displayed.}
    \label{fig:spg-diff_clusters}
\end{figure} 

\begin{figure}[t]
    \centering
    \includegraphics[width=.45\textwidth]{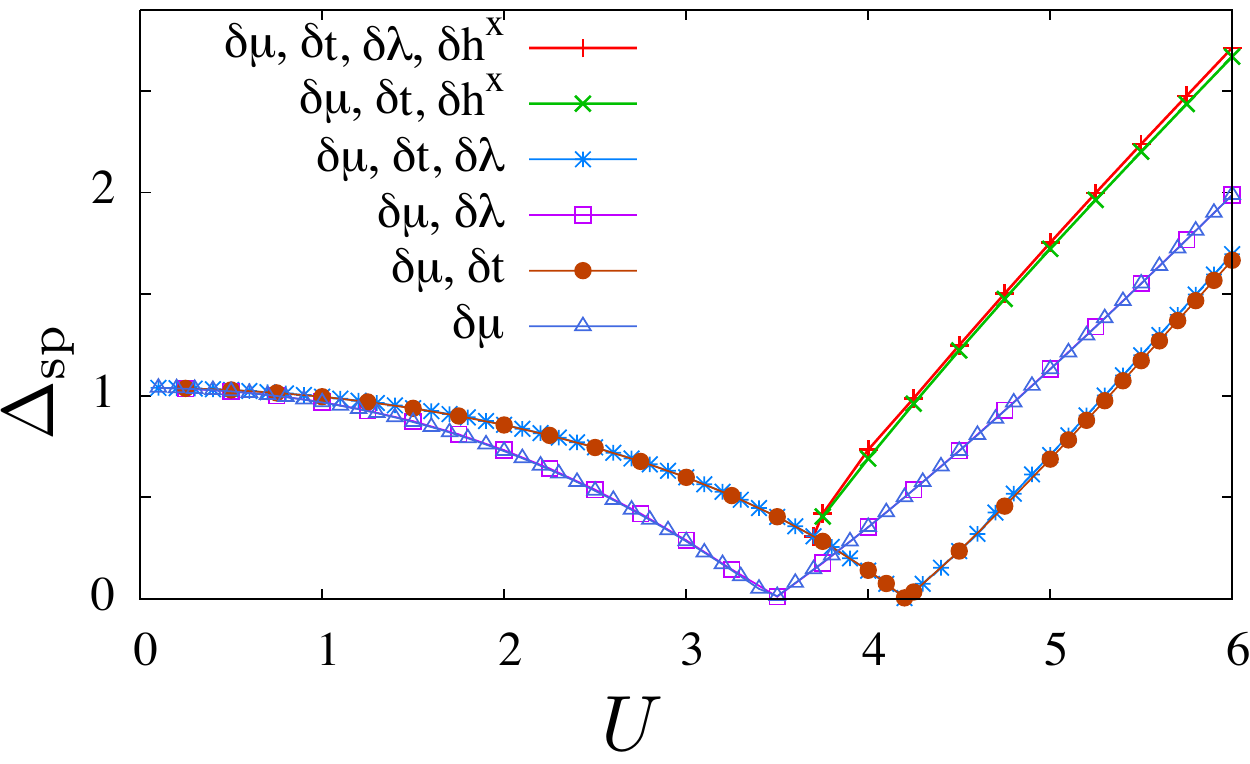} 
    \caption{(Color online). Single-particle gap $\Delta_{\rm sp}$ as a function of $U$ at $\lambda=0.1$ for the six-site cluster. Different combinations of single-particle parameters ($\delta\lambda$, $\delta\mu$, $\delta t$, and Weiss fields $\delta h^x$) are varied to yield a saddle point solution of the grand potential $\Omega$. 
Varying $\delta h^x$, one can see that the single particle gap does not close at the phase transition between the TI and the XY-AFM phase (red and green curves).}
    \label{fig:spg-6site}
    \end{figure}

\begin{figure}[t]
    \centering
    \includegraphics[width=.45\textwidth]{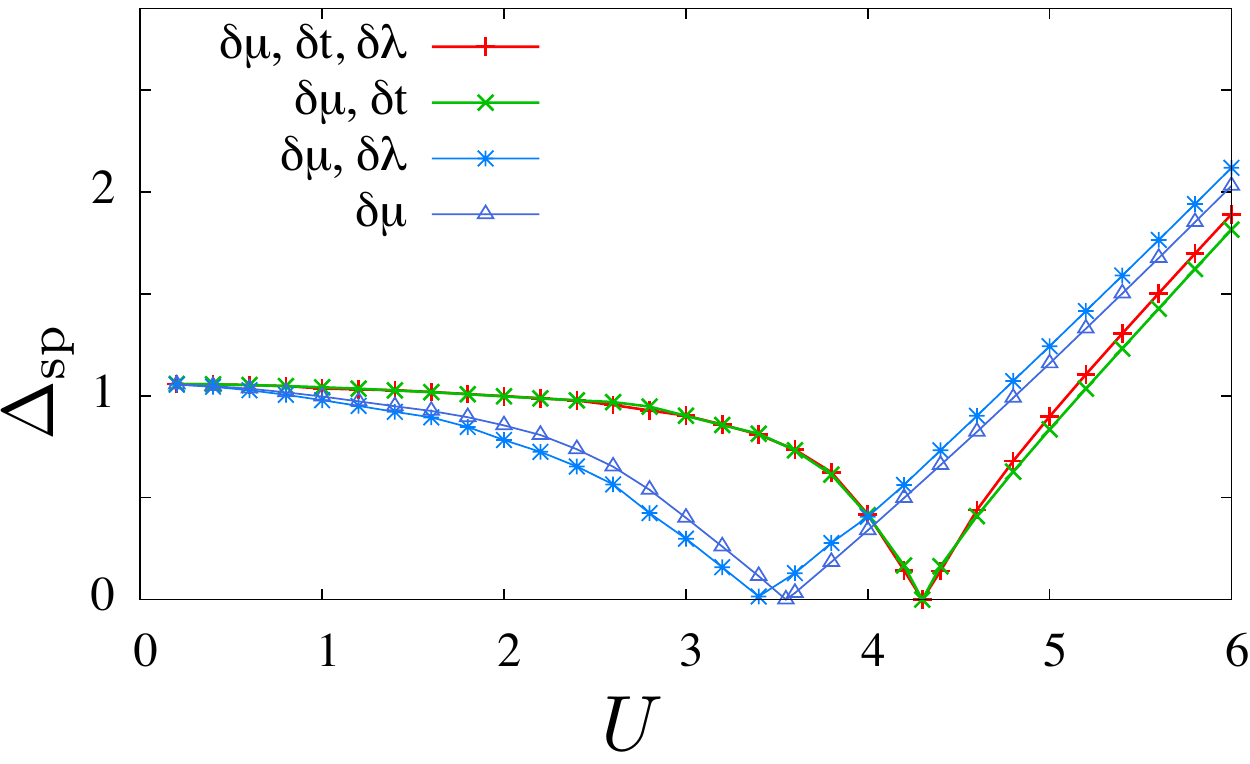} 
    \caption{(Color online). Single-particle gap $\Delta_{\rm sp}$ as a function of $U$ at $\lambda=0.1$ for the eight-site cluster analogous to Fig.~\ref{fig:spg-6site}.}
    \label{fig:spg-8site}
    \end{figure}

The same issue was recently addressed by Liebsch and Wu\,\cite{Liebsch-13prb205127} and also by Hassan and Senechal\,\cite{hassan-13prl096402}. There, it is argued that one bath site per correlated cluster site is not sufficient; at least two bath sites per cluster site should be taken into account\,\cite{hassan-13prl096402}. Liebsch and Wu disagreed and attributed the opening of the single particle gap in case of the ring-shaped six-site cluster only to the geometry of the cluster and the breaking of translational symmetry in methods such as VCA\,\cite{Liebsch-13prb205127}. We confirm in our analysis that the breaking of translational symmetry is problematic, if not detrimental, for a semi-metal state; we will explain below, however, that the breaking of translational symmetry affects other clusters as well which do not possess the six-fold rotational symmetry of the six-site cluster. In any case, both Ref.\,\onlinecite{hassan-13prl096402} and Ref.\,\onlinecite{Liebsch-13prb205127} agree that the opening of the single particle-gap for infinitesimal $U$, as seen for the six-site cluster, is a numerical artifact of the approach and not physically relevant. Inspired by Ref.\,\onlinecite{Liebsch-13prb205127} we plot the single-particle gap as a function of $U$ ($\lambda=0$) for various different clusters (Fig.\,\ref{fig:spg-diff_clusters}). 
As the main result we observe that the semi-metallic phase is never stable with respect to $U$ for the six-site cluster. 

In contrast, the eight- and ten-site clusters seem to provide a stable semi-metallic phase up to finite $U_c$, which we now study in more detail. None of these clusters exhibit the rotational symmetry of the honeycomb lattice. The eight- and the ten-site clusters consists of a single hexagon with two additional ``legs'' on opposite sites and two hexagons located next to each other, respectively (Fig.\,\ref{fig:clusters}\,(b) and (c)). 
We calculated the band structure with an increased unit cell corresponding to the eight site cluster. This allows us to take into account the anisotropy. We find that the semi-metallic phase present in the isotropic case persists for weak anisotropies. To be more specific, it turns out that the gap does not open, the position of the Dirac cones moves, however, away from the $\bs{K}$ and $\bs{K}'$ points. (This is understandable, as the 3-fold discrete rotation symmetry protects the position of the Dirac cones in momentum space.)   A rather large anisotropy is required to merge the Dirac cones and gap them out. The situation here is reminiscent of the $t_1$--$t_2$ model on the honeycomb lattice where a similar behavior is known\,\cite{hasegawa-06prb033413}. 
Performing a VCA analysis for the eight-site cluster, we find that the semi-metallic phase of graphene persists up to $U=2.4$. We also observe within VCA, that the position of the Dirac cones is not at $\bs{K}$ or $\bs{K}'$ anymore in agreement with the anisotropic band structure calculation discussed previously ($\bs{K}^{(')}$ refers to the positions of the Dirac cones at $U=0$). The phase diagram with additional SO coupling is presented in Fig.\,\ref{fig:kmh-phasedias}\,(e).
A similar analysis for the ten-site cluster leads to the same conclusions as for the eight-site cluster, see Fig.\,\ref{fig:kmh-phasedias}\,(c)). 
Quantitatively, we find a slightly larger $U_c=2.9$ where the semi metal to N\'eel-AFM transition occurs (Fig.\,\ref{fig:kmh-phasedias}\,(f).)

\subsection{Magnetic transition}

Our findings indicate that the symmetric six-site cluster has the smallest tendency towards the formation of magnetic order. The less symmetric eight-site cluster, in contrast, is significantly more sensitive towards formation of magnetic order and thus underestimates $U_c$. This is intuitively clear since the eight-site cluster exhibits two ``open legs'', \ie links which have an {\it end site}. These end sites are particularly sensitive towards the formation of magnetic order. Ring-shaped clusters such as six- or ten-site clusters, \ie clusters without end sites, require stronger interactions to acquire magnetic order.

Interestingly, we find that the six-site cluster, while inappropriate for the study of the semi-metal phase, is a good choice in order to study magnetism. For the eight site cluster we can draw the opposite conclusion. The ten-site cluster might be an acceptable compromise; it turns out, however, that for the study with Rashba SO coupling also the ten-site cluster is problematic regarding the investigation of magnetism (see Appendix B for details).

\subsection{Variation of single-particle parameters}
We briefly discuss the influence of the variation of different single-particle parameters within the VCA. In principle, any single-particle parameter (\ie $\delta\mu$, $\delta t$, $\delta\lambda$, $\delta\lambda_R$) can, and should, be varied.
Note that the actual value of a single-particle parameter is, \eg $\mu+\delta\mu$, where $\mu$ is the chosen parameter and $\delta\mu$ comes from the variational scheme.
For practical purposes, however, the variation is often restricted to the variation of $\delta\mu$ only. It is then argued that the additional variation of other single-particle parameters does not affect the results anymore. 
For the six-site cluster, we have already shown for $\lambda=0$ in Fig.\,\ref{fig:spg-diff_clusters}, that the additional variation of $\delta t$ changes the $\Delta_{\rm sp}$-curve quantitatively. We also studied this influence for the TI phase at $\lambda=0.1$ for six- and eight-site clusters. In Fig.\,\ref{fig:spg-6site} the single particle gap $\Delta_{\rm sp}$ of the six-site cluster is shown for the case where (i) $\delta\mu$ only is varied
 (dark-blue curve), (ii) $\delta\mu$ and $\delta t$ are varied (dark-red curve), (iii) $\delta \mu$ and $\delta\lambda$ are varied (pink), (iv) $\delta\mu$, $\delta t$, and $\delta \lambda$ are varied (light-blue). 
Additional variation of the Weiss field $\delta h^x$ is also considered for case (ii) and (iv) (green and red), which reveals that the single-particle gap is not closing at the transition between the TI and the XY-AFM phase\,\cite{rachel13arXiv:1310.3159,wu-12prb205102}, in agreement with QMC results\,\cite{hohenadler-11prl100403}.
 
 \begin{figure}[t!]
\centering
\includegraphics[scale=0.6]{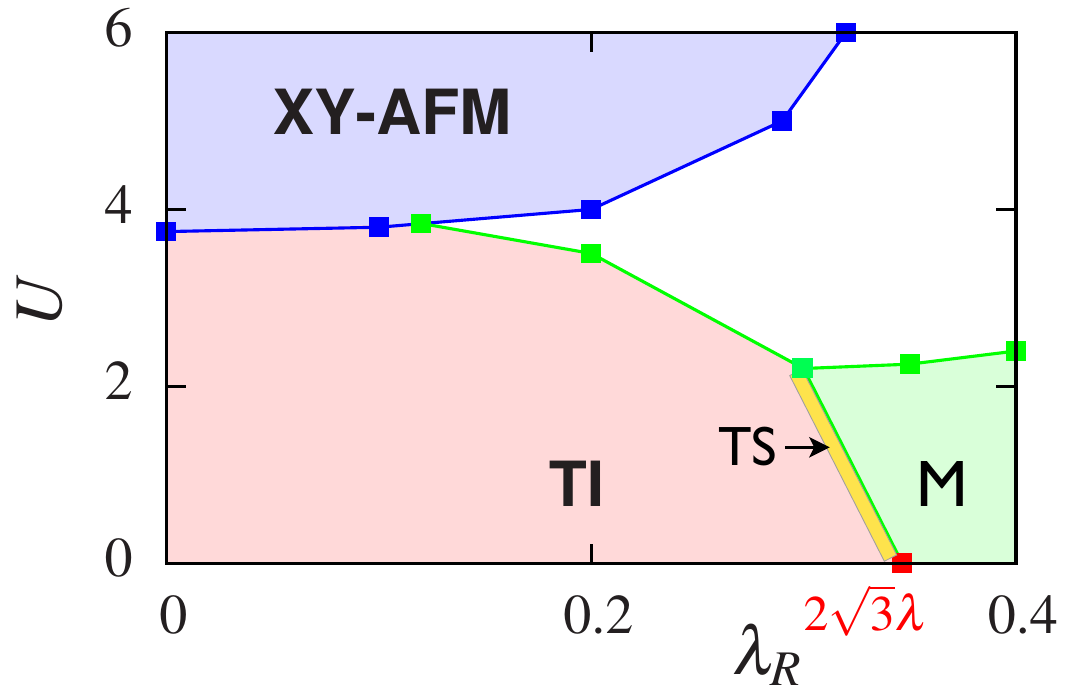}
\caption{(Color online). $U$-$\lambda_R$ KMH phase diagram for $\lambda=0.1$ obtained for an eight-site cluster. In the weak-$\lambda_R$ region, only TI and XY-AFM phases exist. 
The topological semiconductor (TS) phase is very small for $\lambda=0.1$, but increases with $\lambda$. At larger $\lambda_R$ the system is in a metallic phase. In the regime of large $U$ and large $\lambda_R$ no magnetic solution commensurate with the eight-site cluster is found.}
\label{fig:phasedia_lso01_rashba}
\end{figure}

\begin{figure*}
        \begin{center}
            \includegraphics[width=0.97\textwidth]{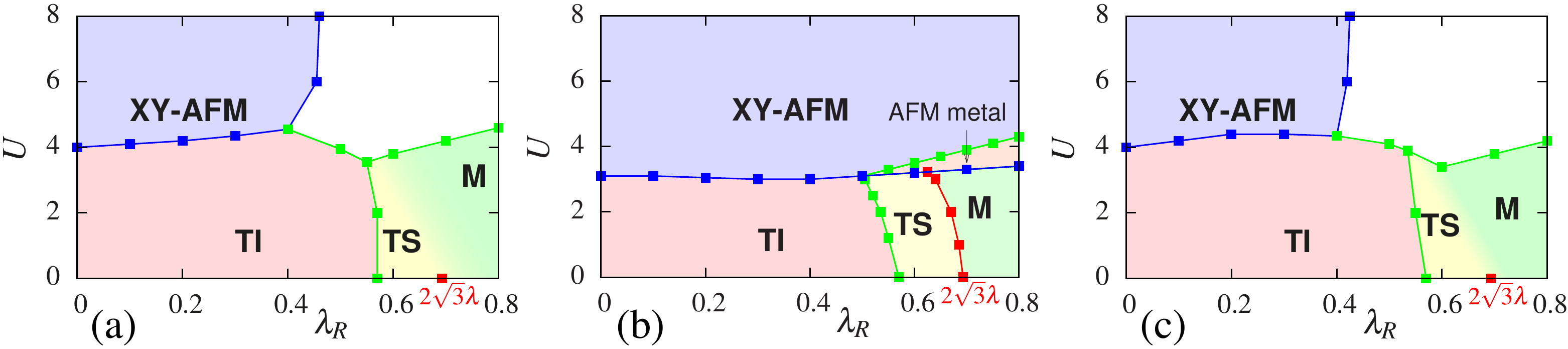}
        \end{center}
    \caption{(Color online). $U$--$\lambda_R$ KMH phase diagram for $\lambda=0.2$ using the (a) six-site cluster, (b) eight-site cluster, and (c) ten-site cluster. Besides the TI and XY-AFM phase, we find a metal (M) phase (green) and a topological semiconductor (TS) phase (yellow) which is characterized by the joint occurrence of helical edge states and zero indirect bulk gap. The  topological-to-metal phase transition for $U=0$ takes place at $\lambda_{R}=2\sqrt{3}\lambda$ (yellow to green phase). The green boundary is obtained by checking whether (i) the bulk gap is closed (LDOS) and whether (ii) edge states are present. At the red boundary, the edge states eventually vanish and one enters a conventional metallic state. For the six- and ten-site clusters we do not find magnetic solutions for $\lambda_R > 0.4$. For the eight-site cluster, we still find N\'eel order and an antiferromagnetic metal state characterized by magnetic order and a zero indirect bulk gap (see also Fig.~\ref{fig:afm-metal}).}
\label{fig:phasedia_lso02_rashba}
\end{figure*}

Essentially, we find that the additional variation of $\delta t$ is important and has significant effects, which also applies to parameter regimes at finite $\lambda$. It should, hence, be generally taken into account in the variational scheme. The additional variation of $\delta\lambda$, however, might lead to new stationary points but can be neglected as it has only negligible effects (Fig.\,\ref{fig:spg-8site}). 
The same conclusion can be drawn for $\delta\lambda_R$.
Since the effect of additional variation of $t$ affects all the phases and all the cluster shapes, we find that at least on the honeycomb lattice, one should always vary $\delta\mu$ and $\delta t$ to obtain reliable VCA results.

\section{Cluster analysis of the KMH model ($\bs{\lambda_R>0}$)}

\subsection{Cluster-dependence of the phase diagram}

In Fig.\,\ref{fig:phasedia_lso01_rashba} we show the phase diagram for the eight-site cluster at $\lambda=0.1$. For this parameter, the TS phase is extremely small and very difficult to detect. Therefore we consider larger intrinsic SO coupling.
Fig.\,\ref{fig:phasedia_lso02_rashba} displays the phase diagrams for the six-, eight-, and ten-site clusters at $\lambda=0.2$. Only for the eight-site cluster (middle panel) we computed edge states which allows us to determine the phase boundary between the TS phase and the metal (red squares). 
Note that we could likewise perform the analogous computation for armchair edges in the case of six- and ten-site clusters. We do not expect, however, further insights from such an additional computation.

\begin{figure}[t]
\centering
\includegraphics[scale=1.2]{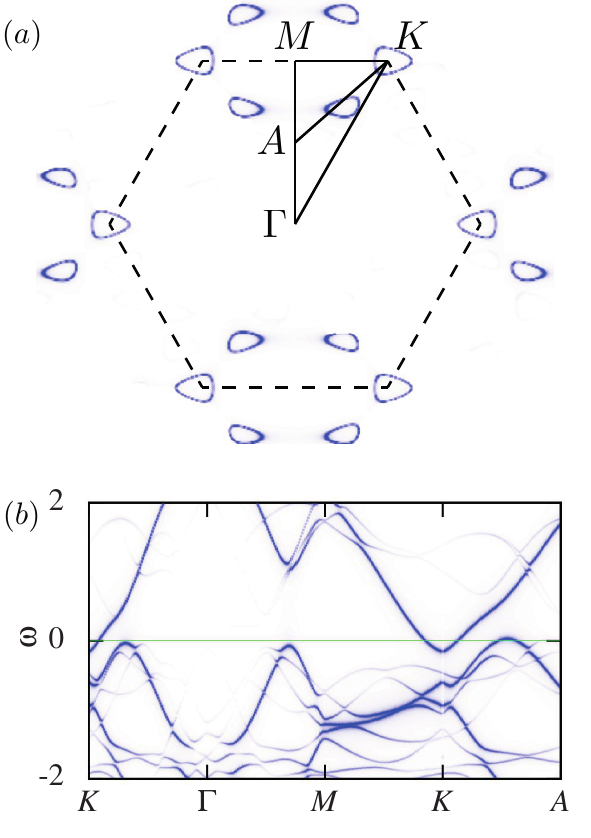}
\caption{(Color online). (a) Fermi surface in the AFM metal phase ($\lambda=0.2$, $\lambda_R=0.6$, and $U=3.3$). (b) Single particle spectral function $A(\bs{k},\omega)$ in the AFM metal phase for periodic boundary conditions, plotted along the trajectory shown in (a).}
\label{fig:afm-metal}
\end{figure}

For the eight- and ten-site clusters, calculating the magnetic domain for strong interactions is different from the six-site cluster. The Rashba term acts differently on different links since it depends on $\bs{\sigma}\times\bs{d}$. Consequently, the results also depend on the orientation of the cluster. The three different nearest-neighbor links of the honeycomb lattice  $\bs{\delta}_1$, $\bs{\delta}_2$, and $\bs{\delta}_3$ are shown in Fig.\,\ref{fig:soc}. 
It is obvious that a cluster (\eg the eight-site cluster) which consists of different numbers of $\bs{\delta}_1$, $\bs{\delta}_2$ and $\bs{\delta}_3$ links, induces a certain anisotropy.
Only the ring-shaped six-site cluster exhibits equal numbers of all $\bs{\delta}_i$- links. Therefore, we should consider the results obtained using the six-site cluster as the most reliable reference. Note, however, that we also incorporated the results for eight- and ten-site clusters and eventually argue that the semi-quantitative phase diagram should look like Fig.\,\ref{fig:schematic-phasedia}.

\begin{figure}[b]
    \begin{center}
        \includegraphics[width=.35\textwidth]{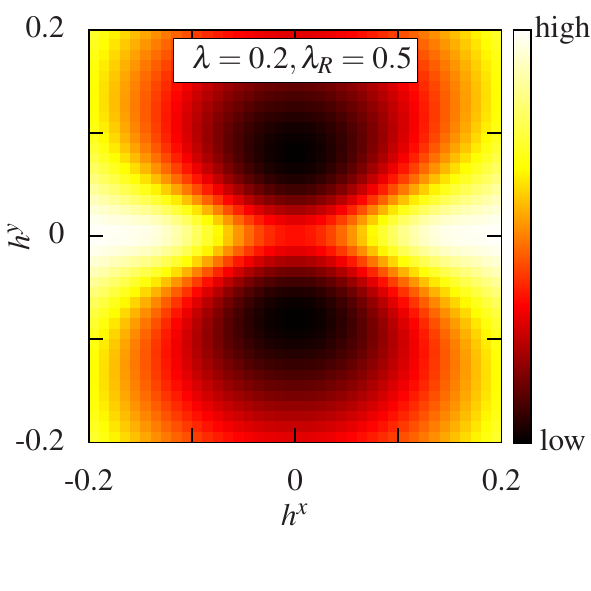} 
    \end{center}
    \caption{Grand potential heat map as a function of antiferromagnetic Weiss fields, $\Omega(h^x,h^y)$ for $\lambda=0.2,\lambda_{R}=0.5$, and $U=4$ on the eight-site cluster. Due to cluster anisotropy, the magnetization points in $y$-direction.}
    \label{fig:af-xy-so-km02-ra05}
\end{figure}

\subsection{AFM metal phase and magnetism}

For the eight-site cluster, another interesting  situation arises. Even for strong $\lambda_R$ and $U$, we find XY-AFM order (for $\lambda=0.1$ and $0.2$). For $\lambda_R>0.5$ and $\lambda=0.2$, however, there is a narrow intermediate-$U$ phase which is an {\it antiferromagnetic metal}. Similar to the topological semiconductor (TS) phase, the strong Rashba coupling bends the bands and gives rise to a metallic density of states. Locally (in momentum space) there is always a direct gap for each wave vector $\bs{k}$. In contrast to the TS phase, there are no edge states but instead a finite magnetization; thus we shall call the phase an antiferromagnetic metal. To provide a better understanding  of this phase, we show in Fig.\,\ref{fig:afm-metal} the bulk spectral function $A(\bs{k}, \omega)$ along the path $K\to\Gamma\to M \to K\to A$. In this plot, one can easily observe that the system is globally gapless, but locally in momentum space there is always a direct gap for each wave vector $\bs{k}$.
We stress that the eight-site cluster exhibits some bias to support such a phase since the onset of magnetization appears for weaker $U$ as compared to other clusters (Fig.\,\ref{fig:phasedia_lso02_rashba}). 

We further find that the antiferromagnetic order loses its U(1) rotation symmetry in the $xy$ plane. We attribute this effect to the different numbers of $\bs{\delta}_1$-, {$\bs{\delta}_2$-} and $\bs{\delta}_3$-bonds in the eight-site cluster, which induces anisotropies when Rashba coupling is present.
In Fig.\,\ref{fig:af-xy-so-km02-ra05} we show the grand potential $\Omega$ as a function of $h^x$ and $h^y$, indicating an antiferromagnetic state pointing in the $y$-direction.
We emphasize, however, that changing the orientation of the eight-site cluster also rotates the direction of the antiferromagnetic order. This shows that anisotropies in the $xy$ plane are cluster artifacts. We, hence, conclude that the actual magnetic order is of XY-AFM type. For larger Rashba coupling, we still find magnetic solutions using the eight-site cluster (\eg the XY-AFM persists up to $\lambda_R \sim 1.36$ at $U=8$).

The ten-site cluster likewise contains 
different numbers of $\bs{\delta}_i$-links, leading to similar anisotropies as for the eight-site cluster.
Around $\lambda_R\sim 0.4$ we observe a breakdown of the magnetic phase (compatible with the results for the six-site cluster). Therefore, we conclude that the resulting VCA phase diagram does not exhibit a magnetically ordered phase for large $\lambda_R$ and large $U$ which would be consistent with a magnetic unit cell provided by the small cluster. The aforementioned AFM metal phase, not present for the ten-site cluster, is most likely an artifact of the eight-site cluster and hence omitted from the final phase diagram in Fig.\,\ref{fig:schematic-phasedia}.

\bibliographystyle{prsty}
\bibliography{kmrh}

\end{document}